\documentclass[prd,showpacs,preprint, 11 pt]{revtex4}
\usepackage{graphicx,color}
\begin{document}

\title{FCNC-induced semileptonic decays of   $J/\psi$
in the Standard Model}

\author{Yu-Ming Wang$^{1}$} \author{Hao Zou$^{1}$} \author{Zheng-Tao Wei$^{2}$}
\author{Xue-Qian Li$^{2}$} \author{Cai-Dian L\"{u} $^{1}$}


 \affiliation{$^{1}$Institute of High Energy Physics, P.O. Box
 918(4), Beijing 100049, China}

 \affiliation{$^{2}$Department of Physics, Nankai University, Tianjin
 300071, China}

\vspace*{1.0cm}

\date{\today}
\begin{abstract}
In this work, we calculate the form factors for $J/\psi\to
\bar{D}^{(*)0}$ induced by the flavor changing neutral currents
(FCNC) in terms of the QCD sum rules. Making use of these form
factors, we further calculate the branching fractions of
semileptonic decays $J/\psi \to \bar{D}^{(*)0} l ^+ l^-$ ($l=e,
\mu$). In particular, we formulate the matrix element $\langle
J/\psi|T_{\mu\nu}| \bar{D}^{*0}\rangle$ with $T_{\mu\nu}$ being a
tensor current, which was not fully discussed in previous
literature. Our analysis indicates that if only the standard model
(SM) applies, the production of single charmed mesons at the present
electron-positron colliders is too small to be observed even the
resonance effects are included, therefore if an anomalous production
rates are observed, it would be a hint of new physics beyond SM.
Even though the predicted branching ratios are beyond the reach of
present facilities which can be seen from a rough order estimate,
the more accurate formulation of the three point correlation
function derived in this work has theoretical significance and the
technique can also be applied to other places. In analog to some
complicated theoretical derivations which do not have immediate
phenomenological application yet, if the future experiments can
provide sufficient luminosity and accuracy, the results would be
helpful.

\end{abstract}

\pacs{13.20.Gd, 14.65.Dw, 11.55.Hx} \maketitle

\newpage

\section{Introduction}

It is widely considered that rare decays of $J/\psi$ can offer an
ideal opportunity to study non-perturbative QCD effects and the
underlying dynamics \cite{Sanchis-Lonzano,Li:2008ey}. On other
aspect, compared with the extensive studies of strong and
electromagnetic decays of $J/\psi$, both experimental and
theoretical investigations of weak decays of $J/\psi$ are much
behind due to their small fractions.

Thanks to the progress in accelerator and detector techniques,
interest in the weak decays of charmonium is being resurgent. With
incomparably large database on $J/\psi$ and other $\psi-$family
members, the BES collaboration will measure some rare decays
including the semi-leptonic \cite{BES} and non-leptonic modes
\cite{BESIII:2007mr} with high accuracy, and more further
theoretical and experimental studies would follow. Theoretically,
the semileptonic decays of $J/\psi$ induced by the flavor changing
currents  were analyzed in our earlier work \cite{Wang:2007ys},
where the QCD sum rules (QCDSR) approach
\cite{Shifman:1978bw,Shifman:1978bx,Shifman:1978by,Novikov:1981xj,reinders}
was employed to compute the transition form factors. Subsequently,
by utilizing  the form factors obtained in terms of  QCDSR we
carried out computations on the rates of non-leptonic decays of
$J/\psi$ \cite{Wang:2008tf} under the factorization assumption. Very
recently, weak decay of $J/\psi$ into the final states involving a
pseudoscalar meson were also studied by authors of Ref.
\cite{shenyl} where the covariant light front quark model was
employed, thus their result can be regarded as a cross check of that
estimated in QCDSR.

At the quark level, the  decay of $J/\psi$ induced by the
flavor-changing neutral current (FCNC) is realized via $c \to u$
transition, which should be very small due to the
Glashow-Iliopoulos-Maiani (GIM) mechanism \cite{Glashow:1970gm},
whereas at the hadron level the long-distance effects may have the
same order of magnitude. Although the FCNC processes for $B$ and $K$
cases are comprehensively studied, the FCNC decays in the charmed
mesons has not caught enough attention due to the stronger GIM
suppression for up-type quarks, which is also responsible for
smallness of $D^0-\bar{D}^0$ mixing
\cite{Ohl:1992sr,Donoghue:1985hh,Petrov:1997ch,Golowich:1998pz,Falk:2004wg}.
As aforementioned, the progress of detection techniques and
facilities allows much more accurate measurements on the rare
decays, so theoretically we need to calculate the production rates
and see if the expected precision is indeed possible to observe a
non-zero fraction at the updated facilities. Thus, in this work, we
would like to take a step forward to investigate the FCNC processes
$J/\psi \to \bar{D}^{(*)0} l^{+}l^{-}$ in the standard model.

Following the procedure given in Ref.
\cite{Wang:2007ys,Wang:2008tf}, we will employ the three-point QCDSR
to derive the form factors. The QCDSR has  been proved to be an
effective tool to calculate various hadronic matrix elements where
non-perturbative QCD effects dominate. The sum rule technique for
three-point correlation functions  was first used to describe the
pion electromagnetic form factor at intermediate momentum transfer
\cite{ioffe 1, nesterenko} and hence this approach has been applied
to various weak decays \cite{weak decays of QCDSR 1, weak decays of
QCDSR 2}. An alternative approach is the light-cone sum rules where
the light-cone distribution amplitudes of hadrons are employed
\cite{Chernyak,wf of baryons, chenghyscalar, scalarwf} to calculate
the form factors in similar processes.  In this work, we only
concentrate ourselves in the QCDSR. To evaluate a transition
process, calculation of three-point correlation function is needed
and obviously it is much more complicated than the calculations of
two-point correlations.

The structure of this paper is organized as follows: After this
introduction, we will firstly display the effective Hamiltonian
relevant to the semileptonic decay $J/\psi\to \bar{D}^{(*)0}$ and
then derive the sum rules for the form factors in section \ref{The
standard procedure}. The Wilson coefficients of various operators
contributing to the correlation functions  are calculated in much
detail in section \ref{Wilson coefficients} making use of the
operator product expansion technique. In particular, the Wilson
coefficients of gluon condensate and quark gluon mixing operator are
dealt with in the fixed-point gauge, i.e., $x_{\mu}A^{a}_{\mu}=0$.
Furthermore, the inputs for the numerical computations of form
factors are presented at the beginning of section \ref{Numerical
results}, and then an extensive analysis of sum rules of the form
factors are performed. We explicitly show the Borel platform where
the form factors are stable with respect to variations of the Borel
masses $M_1$ and $M_2$. The rates of semileptonic  decays $J/\psi$
to $\bar{D}^{(*)0}$ are numerically evaluated in section \ref{decay
rate}, and the last section is devoted to our discussions and
conclusions.

\section{The standard procedure}
\label{The standard procedure}
\subsection{Effective Hamiltonian for semileptonic decays of $J/\psi$ to  $\bar{D}^{(*)0}$}
The quark level FCNC transition $c \to u l^{+} l^{-}$ for the
semileptonic decay of $J/\psi \to \bar{D}^{(*)0} l^{+} l^{-} $ is
described by the effective Hamiltonian
\begin{eqnarray}
\mathcal{H}_{eff}(c \to u)&=&-{G_{F} \over 4 \sqrt{2}}{\alpha_{em}
\over \pi} [C_9^{eff}(\mu) \bar{u}\gamma_{\mu}(1-\gamma_5)c
\bar{l}\gamma^{\mu}l+C_{10}(\mu) \bar{u}\gamma_{\mu}(1-\gamma_5)c
\bar{l}\gamma^{\mu}\gamma_5 l \nonumber \\
&&-2m_c C_7^{eff}(\mu)\bar{u}i\sigma_{\mu \nu}{q^{\nu} \over
q^2}(1+\gamma_5)c\bar{l}\gamma^{\mu}l], \label{effective lagrangian}
\end{eqnarray}
with $q$ being the momentum of the lepton pair. In Eq.
(\ref{effective lagrangian}), the Wilson coefficients contain the
Cabibbo-Kobayashi-Maskawa (CKM) matrix elements. The explicit forms
of $C_{7,9}^{eff}(\mu)$ and $C_{10}$ can be found in literature
\cite{buras, golowich 1, golowich 2, greub, aliev} which are
displayed as follows
\begin{eqnarray}
C_9^{eff}(\mu)=C_{9}(\mu)+C_9^{con}(z_q,s,\mu)+C_9^{res}(z_q,s,\mu), \nonumber \\
C_7^{eff}(\mu)=C_{7}(\mu)+C_7^{con}(z_q,s,\mu)+C_7^{res}(z_q,s,\mu),
\end{eqnarray}
where the functions $C_i^{con}(z_q,s,\mu)$ and
$C_i^{res}(z_q,s,\mu)$ represent the contributions from the
continuum and resonance parts of self-energy loops of $d \bar{d}$,
$s \bar{s}$ and $b \bar{b}$ and $z_q$ and $s$ are defined as
$z_q=m_{q} / m_c$, $s= q^2/m_c^2$ with the subscript $q$ denoting
$d$, $s$ and $b$ quarks. $C_9^{con}(z_q,s,\mu)$ caused by the
leading order mixing between $O_1$ with $O_9$ is given
as\cite{golowich 2}
\begin{eqnarray}
C_9^{con}(z_q,s)=\sum_{q=d,s,b} \lambda_{q} [-{2 \over 9}
{\rm{ln}}{m_q^2 \over M_W^2}+{8 \over 9}{z_{q}^2 \over s}-{1 \over
9}(2+{4 z_{q}^2 \over s})\sqrt{|1-{4 z_{q}^2 \over s}|}\,\, T(z_q)]
\end{eqnarray}
with
\begin{eqnarray}
T(z_q)=
 \left\{
\begin{array}{l}
2 {\rm{arccot}} (\sqrt{{4 z_{q}^2 \over s}-1}), \qquad
{{\rm{for}} \,\, s<4 z_q^2}; \\
{\rm{ln}} \bigg| {{1+\sqrt{1-{4 z_{q}^2 \over s}}} \over
{1-\sqrt{1-{4 z_{q}^2 \over s}}}}\bigg| -i \pi, \qquad {{\rm{for}}
\,\, s>4 z_q^2},
\end{array}
\right.
\end{eqnarray}
where $\lambda_q$ is the CKM matrix element
$\lambda_q=V_{cq}^{*}V_{uq}$. The contributions of resonances from
quark loops to $C_9^{eff}(\mu)$ can be expressed by
$C_9^{res}(z_q,s,\mu)$ as a shift of the Wilson coefficient
$C_9(\mu)$.  $C_9^{res}(z_q,s,\mu)$ is given in \cite{golowich 2}
\begin{eqnarray}
C_9^{res}(z,s)={3 \pi^2 \over \alpha_{em}^2} \sum_{i} \kappa_{i}
{m_{V_i}\Gamma_{V_i \to l^{+}l^{-}} \over m_{V_i}^2 -q^2- i
m_{V_i}\Gamma_{V_i}},
\end{eqnarray}
where $\kappa_{i}$ is a free parameter to compensate the deviation
caused by the approximation of native factorization
\cite{ali,lim,neubert}, and can be adjusted to reproduce the
branching ratio of non-leptonic decays $D \to V_{i} X$. The numbers
of $\kappa_i$ for light vector mesons were calculated in Ref.
\cite{golowich 2} as $\kappa_{\rho}=0.7$, $\kappa_{\omega}=3.1$ and
$\kappa_{\phi}=3.6$. $C_7^{con}(z_q,s,\mu)$ and
$C_7^{res}(z_q,s,\mu)$ are small and can be neglected.

For readers' convenience, we collect the Wilson coefficients at
$\mu=m_W$ as
\begin{eqnarray}
C_7(m_W)&=&-\sum_{q=d,s,b}\lambda_{q} F_2(x_q), \nonumber \\
C_9(m_W)&=&{1 \over {\rm{sin}}^2\Theta_W} \sum_{q=d,s,b} \lambda_{q}
[(C^{box}(x_q) +C^Z(x_q))-2{\rm{sin}}^2\Theta_W
(F_1(x_q)+C^Z(x_q))],
\nonumber \\
C_{10}(m_W)&=&-{1 \over {\rm{sin}}^2\Theta_W} \sum_{q=d,s,b}
\lambda_{q}(C^{box}(x_q) +C^Z(x_q)),
\end{eqnarray}
where $x_q=m_q^2/m_W^2$, $\Theta_W$ is the weak mixing angle. The
explicit expressions for $F_1(x_q), F_2(x_q), C^{box}(x_q)$ and
$C^Z(x_q)$ can be found in Ref.\cite{inami,aliev} and are also
included in our Appendix \ref{FCNC function}.

In order to obtain the decay rates of $J/\psi\to\bar{D}^{(*)0}$, we
need to calculate the hadronic matrix elements which are usually
parameterized in the following forms \cite{wirbel,p. ball
parameterization, beneke}:
\begin{eqnarray}
&&\langle \bar{D}^0(p_2)|\bar{q} \sigma_{\mu
\nu}q^{\nu}(1+\gamma_5)c|J/\psi(\epsilon,p_1)\rangle\nonumber \\
&&~~~~=-2i\epsilon_{\mu \rho \alpha
\beta}\epsilon^{\rho}p_1^{\alpha}p_2^{\beta}T_1(q^2)-[\epsilon_{\mu}(m_{\psi}^2-m_{D}^2)-(\epsilon
\cdot q) (p_1+p_2)_{\mu}]T_2(q^2)\nonumber \\
&&~~~~~~~~+(\epsilon \cdot q)[q_{\mu}-{q^2 \over
m_{\psi}^2-m_{D}^2}(p_1+p_2)_{\mu}]T_3(q^2),
\\
&&\langle \bar{D}^{*0}(\epsilon_2,p_2)|\bar{q} \sigma_{\mu
\nu}q^{\nu}(1+\gamma_5)c|J/\psi(\epsilon_1,p_1)\rangle\nonumber \\
&&~~~~=(m_{\psi}+m_{D^{*}})\epsilon_{\mu \rho \alpha \beta }q^{\rho}
\epsilon_1^{\alpha} \epsilon_2^{*\beta}
\tilde{T}_1(q^2)\nonumber \\
&&~~~~~~~~+{1 \over m_{\psi}^2-m_{D^{*}}^2}\epsilon_{\mu \nu \alpha
\beta} p_1^{\alpha} p_2^{\beta} [\tilde{T}_2(q^2) \epsilon_1^{\nu}
\epsilon_2^{*} \cdot q+\tilde{T}_3(q^2) \epsilon_2^{*\nu} \epsilon_1
\cdot q]  \nonumber\\
&&~~~~~~~~-i (m_{\psi}+m_{D^{*}}) (\epsilon_1 \cdot
\epsilon_2^{*})[{p_1}_{\mu}-{m_{\psi}^2-m_{D^{*}}^2+q^2 \over
m_{\psi}^2-m_{D^{*}}^2-q^2}{p_2}_{\mu}]\tilde{T}_4(q^2)\nonumber \\
&&~~~~~~~~-{i \over m_{\psi}-m_{D^{*}}}(\epsilon_1 \cdot
q)(\epsilon_2^{*} \cdot q)[{p_1}_{\mu}-{m_{\psi}^2-m_{D^{*}}^2+q^2
\over
m_{\psi}^2-m_{D^{*}}^2-q^2}{p_2}_{\mu}]\tilde{T}_5(q^2)\nonumber \\
&&~~~~~~~~-i(m_{\psi}+m_{D^{*}}) [{\epsilon_1}_{\mu}(\epsilon_2^{*}
\cdot
q)-{\epsilon_2}^{*}_{\mu}(\epsilon_1 \cdot q)]\tilde{T}_6(q^2) \nonumber\\
&&~~~~~~~~+{i \over (m_{\psi}-m_{D^{*}})}
[(m_{\psi}^2-m_{D^{*}}^2-q^2) {\epsilon_{1}}_{\mu}-2(\epsilon_1
\cdot q){p_2}_{\mu}] (\epsilon_2^{*} \cdot q) \tilde{T}_7(q^2),
\label{vector tensor}
\end{eqnarray}
where the totally anti-symmetric tensor is defined as
${\rm{Tr}}[\gamma_{\mu} \gamma_{\nu} \gamma_{\rho} \gamma_{\sigma}
\gamma_5]=4 i \epsilon_{\mu \nu \rho \sigma}$ as a convention
adopted in our work. It is worth emphasizing that the
parametrization of hadronic matrix elements for $J/\psi$ decays to
vector charmed meson, shown in Eq.~(\ref{vector tensor}) is new and
has not ever emerged before. Besides, the above parametrization
forms are also chosen by the requirement that the stable platform
with two Borel variables can be achieved to assure our predictions
credible.

\subsection{Sum rules for transition form factors}

\subsubsection{Sum rules for transition form factors of $J/\psi\to\bar{D}^{0}$}
As for the FCNC process $J/\psi\to \bar{D}^{0}$, both the  ``$V-A$"
current and the tensor operator can contribute to the decay
amplitude. Here the former one can be directly obtained from the
case of $J/\psi$ to $D^{-}_{d,s}$  by exchanging $s$ or $d$ quark
into $u$ quark, however, the latter one has not appeared ever
before, hence we should re-derive the sum rules for the form factors
involved in the hadronic matrix element where the tensor operator is
sandwiched between  $J/\psi$ and $\bar{D}^{0}$ states. Following the
standard procedure, the three-point function is set as
\begin{eqnarray}
\tilde{\Pi}_{\mu \nu}=i^2 \int d^4 x d^4 y e^{-i p_1 \cdot y +i p_2
\cdot x}\langle 0 |j_5^{\bar{D}^{0}}(x)
j_{\mu}(0)j_{\nu}^{J/\psi}(y) | 0 \rangle,
\end{eqnarray}
where the current $j_{\nu}^{J/\psi}(y)=\bar{c}(y) \gamma _{\nu}
c(y)$ represents $J/\psi$ channel;  $j_{\mu}(0)=\bar{u} \sigma_{\mu
\nu }(1+\gamma_5)q^{\nu} c$ describes the weak current for $J/\psi$
to $\bar{D}^{0}$ and $j_{5}^{\bar{D}^{0}}(x)=\bar{c}(x) i \gamma_{5}
u(x)$ denotes the $\bar{D}^{0}$ channel. Inserting two complete sets
of states with the quantum numbers of $J/\psi$ and $\bar{D}^{0}$
mesons simultaneously into the above correlation function, one can
arrive at the hadronic representation of the three-point function as
\begin{eqnarray}
\tilde{\Pi}_{\mu \nu}&=&{f_{\bar{D}^{0}} m_{\bar{D}^{0}}^2\langle
\bar{D}^{0}|j_{\mu}|J/\psi\rangle m_{J/\psi}
f_{J/\psi}\epsilon_{\nu}^{*\lambda} \over
(m_{J/\psi}^2-p_1^2)(m_{\bar{D}^{0}}^2-p_2^2)(m_c+m_u)}+ \int\int
_{\sum_{12}}ds_1 ds_2 {\tilde{\rho}^{h}_{\mu \nu}(s_1,s_2,q^2) \over
(s_1-p_1)^2(s_2-p_2)^2} \nonumber \\ && +\mathrm{subtraction} \,\,\,
\mathrm{terms}.
\end{eqnarray}
The subtraction terms are polynomials of either $p_1$ or $p_2$,
which will disappear after performing the double Borel
transformation $\hat{\mathcal{B}}_{M_1^2}
\hat{\mathcal{B}}_{M_2^2}$, with
\begin{eqnarray}
\hat{\mathcal{B}}_{M_i^2}=\lim_{\stackrel{-p_i^2,n \to
\infty}{-p_i^2/n=M^2}} \frac{(-p_i^2)^{(n+1)}}{n!}\left(
\frac{d}{dp_i^2}\right)^n.
\end{eqnarray}
Applying the operator product expansion technique to the
$\tilde{\Pi}_{\mu \nu}$ in the deep Euclidean region, we achieve the
expression of this correlation function as
\begin{eqnarray}
\tilde{\Pi}_{\mu \nu}=i \tilde{f}_{0} \epsilon_{\mu \nu \alpha \beta
}p_1^{\alpha}p_2^{\beta}+\tilde{f}_1 {p_1}_{\mu} {p_{1}}_{\nu}
+\tilde{f}_2 {p_2}_{\mu} {p_{2}}_{\nu} +\tilde{f}_3 {p_2}_{\mu}
{p_{1}}_{\nu}+\tilde{f}_4 {p_1}_{\mu} {p_{2}}_{\nu}+\tilde{f}_5
g_{\mu \nu},
\end{eqnarray}
with each coefficient ${\tilde{f}}_{i}$ contributed from both
perturbative part and non-perturbative condensate, i.e.,
\begin{eqnarray}
\tilde{f}_i={\tilde{f}}_i^{pert} {\mathbf{I}} + {\tilde{f}}_{i}^{qq}
\langle \bar{q} q\rangle + {\tilde{f}}_{i}^{GG} \langle G G \rangle
+ {\tilde{f}}_{i}^{qGq} \langle \bar{q} G q\rangle +....
\label{tilde fi expansion}
\end{eqnarray}
Comparing the two different expressions for $\tilde{\Pi}_{\mu \nu}$
calculated in the QCD and hadronic representations and performing
the double Borel transformation on variables $p_1$ and $p_2$, we can
extract the sum rules for the form factors involved in the decay
mode of $J/\psi$ to  $\bar{D}^{0}$ as
\begin{eqnarray}
\label{T1 pseudoscalar}T_1(q^2)&=&{m_c+m_u \over
2m_{\psi}f_{\psi}f_{\bar{D}^{0}}m_{\bar{D}^{0}}^2}e^{m_{\psi}^2 /
M_1^2}e^{m_{\bar{D}^{0}}^2 / M_2^2}M_1^2
M_2^2\hat{\mathcal{B}}\tilde{f}_0,
\\
T_2(q^2)&=&{m_c+m_u \over
(m_{\psi}^2-m_D^2)m_{\psi}f_{\psi}f_{\bar{D}^{0}}m_{\bar{D}^{0}}^2}e^{m_{\psi}^2
/ M_1^2}e^{m_{\bar{D}^{0}}^2 / M_2^2}M_1^2
M_2^2\hat{\mathcal{B}}\tilde{f}_5,
\\
T_3(q^2)&=&-{m_c+m_u \over
2m_{\psi}f_{\psi}f_{\bar{D}^{0}}m_{\bar{D}^{0}}^2}e^{m_{\psi}^2 /
M_1^2}e^{m_{\bar{D}^{0}}^2 / M_2^2}M_1^2
M_2^2\hat{\mathcal{B}}(\tilde{f}_2-\tilde{f}_4). \label{T3
pseudoscalar}
\end{eqnarray}

\subsubsection{Sum rules for transition form factors of $J/\psi\to\bar{D}^{*0}$}
Now we are ready to derive the sum rules for the form factors which
are responsible for the decay channel of $J/\psi\to\bar{D}^{*0}$.
Now the three-point function can be written as
\begin{eqnarray}
\tilde{\Pi}_{\mu \nu \rho}=i^2 \int d^4 x d^4 y e^{-i p_1 \cdot y +i
p_2 \cdot x}\langle 0 |j_{\rho}^{\bar{D}^{*0}}(x)
j_{\mu}(0)j_{\nu}^{J/\psi}(y) | 0 \rangle,
\end{eqnarray}
where the current
$j_{\rho}^{\bar{D}^{*0}}(x)=\bar{c}(x)\gamma_{\rho}u(x)$ describes
the $\bar{D}^{*0}$ channel, and $j_{\nu}^{J/\psi}(y)$, $j_{\mu}(0)$
are the same as that in last subsection.  The matrix element defined
by the ``V-A" operator can be gained directly from the decay of
$J/\psi$ to $D^{*-}_{d,s}$ presented in the previous subsection. On
the one hand, one can write the phenomenological representation of
$\tilde{\Pi}_{\mu \nu \rho}$ at the hadron level as
\begin{eqnarray}
{\tilde{\Pi}}_{\mu \nu \rho}&=&{{m_{\bar{D}^{*0}}}
f_{{\bar{D}^{*0}}} {\epsilon'_{\rho}}^{\lambda'}\langle
\bar{D}^{*0}|j_{\mu}|J/\psi\rangle m_{J/\psi}
f_{J/\psi}\epsilon_{\nu}^{*\lambda} \over
(m_{J/\psi}^2-p_1^2)(m_{\bar{D}^{*0}}^2-p_2^2)}+ \int\int
_{\sum_{12}}ds_1 ds_2 {\tilde{\rho}^{h}_{\mu \nu \rho}(s_1,s_2,q^2)
\over (s_1-p_1^2)(s_2-p_2^2)} \nonumber \\ && +\mathrm{subtraction}
\,\,\, \mathrm{terms}.
\end{eqnarray}
On the other hand, the correlation function $\tilde{\Pi}_{\mu \nu
\rho}$ can be calculated at the quark level as
\begin{eqnarray}
\tilde{\Pi}_{\mu \nu \rho}&=& \tilde{F}_{1} \epsilon_{\mu \nu \alpha
\beta }p_1^{\alpha}p_2^{\beta}{p_1}_{\rho}+ \tilde{F}_{2}
\epsilon_{\mu \nu \alpha \beta }p_1^{\alpha}p_2^{\beta}{p_2}_{\rho}+
\tilde{F}_{3} \epsilon_{\mu \rho \alpha
\beta}p_1^{\alpha}p_2^{\beta}{p_1}_{\nu} +\tilde{F}_{4}
\epsilon_{\mu \rho \alpha \beta}p_1^{\alpha}p_2^{\beta}{p_2}_{\nu}+
\tilde{F}_{5} \epsilon_{\nu \rho \alpha
\beta}p_1^{\alpha}p_2^{\beta}{p_1}_{\mu} \nonumber \\
&& + \tilde{F}_{6} \epsilon_{\nu \rho \alpha \beta
}p_1^{\alpha}p_2^{\beta}{p_2}_{\mu} +i \tilde{F}_{7}g_{\mu
\nu}{p_1}_{\rho}+i \tilde{F}_{8}g_{\mu \rho}{p_1}_{\nu}+i
\tilde{F}_{9}g_{\nu \rho}{p_1}_{\mu} +i \tilde{F}_{10}g_{\mu
\nu}{p_2}_{\rho}+i \tilde{F}_{11}g_{\mu \rho}{p_2}_{\nu}+i
\tilde{F}_{12}g_{\nu \rho}{p_2}_{\mu}\nonumber \\
&&+i \tilde{F}_{13}{p_1}_{\mu}{p_1}_{\nu}{p_1}_{\rho} +i
\tilde{F}_{14}{p_2}_{\mu}{p_2}_{\nu}{p_1}_{\rho}+i
\tilde{F}_{15}{p_1}_{\mu}{p_2}_{\nu}{p_1}_{\rho} +i
\tilde{F}_{16}{p_2}_{\mu}{p_1}_{\nu}{p_1}_{\rho}+i
\tilde{F}_{17}{p_2}_{\mu}{p_2}_{\nu}{p_2}_{\rho} \nonumber \\
&&+i \tilde{F}_{18}{p_1}_{\mu}{p_1}_{\nu}{p_2}_{\rho}+i
\tilde{F}_{19}{p_2}_{\mu}{p_1}_{\nu}{p_2}_{\rho} +i
\tilde{F}_{20}{p_1}_{\mu}{p_2}_{\nu}{p_1}_{\rho},
\end{eqnarray}
where each of the above coefficients $\tilde{F}_{i}$ receives both
perturbative and non-perturbative contributions
\begin{eqnarray}
\tilde{F}_i=\tilde{F}_i^{pert} {\mathbf{I}} + \tilde{F}_{i}^{qq}
\langle \bar{q} q\rangle + \tilde{F}_{i}^{GG} \langle G G \rangle +
\tilde{F}_{i}^{qGq} \langle \bar{q} G q\rangle +.... \label{tilde Fi
expansion}
\end{eqnarray}
Finally, equating the above quark-level and hadron-level forms of
$\tilde{\Pi}_{\mu \nu \rho}$, we  obtain the sum rules of the form
factors as
\begin{eqnarray}
\label{tilde T1 vector}
\tilde{T}_1(q^2)&=&{m_{\bar{D}^{*0}}^4-2(q^2+m_{\psi}^2)m_{\bar{D}^{*0}}^2
+(q^2-m_{\psi}^2)^2\over
2(m_{\psi}+m_{\bar{D}^{*0}})(q^2-m_{\psi}^2+m_{\bar{D}^{*0}}^2)
m_{\psi}f_{\psi}m_{\bar{D}^{*0}}f_{\bar{D}^{*0}}}e^{m_{\psi}^2 /
M_1^2}e^{m_{\bar{D}^{*0}}^2 / M_2^2}M_1^2
M_2^2\hat{\mathcal{B}}\tilde{F}_1,
 \\
\tilde{T}_2(q^2)&=&{m_{\psi}^2 - m_{\bar{D}^{*0}}^2 \over
m_{\psi}f_{\psi}m_{\bar{D}^{*0}}f_{\bar{D}^{*0}}}e^{m_{\psi}^2 /
M_1^2}e^{m_{\bar{D}^{*0}}^2 /
M_2^2}M_1^2 M_2^2\hat{\mathcal{B}}(\tilde{F}_{1}-\tilde{F}_{5}),
\\
\tilde{T}_3(q^2)&=&-{m_{\psi}^2 - m_{\bar{D}^{*0}}^2 \over
(q^2-m_{\psi}^2+m_{\bar{D}^{*0}}^2)m_{\psi}f_{\psi}m_{\bar{D}^{*0}}f_{\bar{D}^{*0}}}e^{m_{\psi}^2
/ M_1^2}e^{m_{\bar{D}^{*0}}^2 / M_2^2}M_1^2 M_2^2 \nonumber \\
&& \times  \hat{\mathcal{B}}[(\tilde{F}_{4}+\tilde{F}_{5})q^2
+(\tilde{F}_{4}-\tilde{F}_{5})(m_{\bar{D}^{*0}}^2-m_{\psi}^2)],
\\
\tilde{T}_4(q^2)&=&-{1 \over
(m_{\psi}+m_{\bar{D}^{*0}})m_{\psi}f_{\psi}m_{\bar{D}^{*0}}f_{\bar{D}^{*0}}}e^{m_{\psi}^2
/ M_1^2}e^{m_{\bar{D}^{*0}}^2 / M_2^2}M_1^2
M_2^2\hat{\mathcal{B}}\tilde{F}_{9},
\\
\tilde{T}_5(q^2)&=&-{m_{\bar{D}^{*0}}-m_{\psi} \over
m_{\psi}f_{\psi}m_{\bar{D}^{*0}}f_{\bar{D}^{*0}}}e^{m_{\psi}^2 /
M_1^2}e^{m_{\bar{D}^{*0}}^2 / M_2^2}M_1^2
M_2^2\hat{\mathcal{B}}\tilde{F}_{15},
\\
\tilde{T}_6(q^2)&=&-{1 \over
(m_{\psi}+m_{\bar{D}^{*0}})m_{\psi}f_{\psi}m_{\bar{D}^{*0}}f_{\bar{D}^{*0}}}e^{m_{\psi}^2
/ M_1^2}e^{m_{\bar{D}^{*0}}^2 / M_2^2}M_1^2
M_2^2\hat{\mathcal{B}}\tilde{F}_{11},
\\
\tilde{T}_7(q^2)&=&{m_{\bar{D}^{*0}}-m_{\psi} \over
(q^2+m_{\bar{D}^{*0}}^2-m_{\psi}^2)m_{\psi}f_{\psi}m_{\bar{D}^{*0}}f_{\bar{D}^{*0}}}e^{m_{\psi}^2
/ M_1^2}e^{m_{\bar{D}^{*0}}^2 / M_2^2}M_1^2
M_2^2\hat{\mathcal{B}}(\tilde{F}_{7}-\tilde{F}_{11}).
 \label{tilde T7 vector}
\end{eqnarray}
Now we have achieved the sum rules for the form factors, the next
step is to calculate the Wilson coefficients corresponding to the
various operators in the operator product expansion at the deep
Euclidean region ($-q^2 \gg 0$) in next section.

\section{The calculations of Wilson coefficients}
\label{Wilson coefficients}

In this section we calculate the Wilson coefficients. To guarantee
sufficient theoretical accuracy, the correlation functions are
required to be expanded up to dimension-5 operators, namely
quark-gluon mixing condensate. The dimension-6 operators, such as
the four quark condensates, are small and further suppressed by
$O({\alpha}_s^2)$, so can be safely neglected in our calculations.

\subsection{Wilson coefficients of the correlation function $\tilde{\Pi}_{\mu \nu}$}
\label{Wilson coefficients 1}

The diagrams which depict the contributions from the perturbative
part and nonperturbative condensates are shown in Fig.~\ref{wilson
coefficients graph}. The first diagram results in the Wilson
coefficient of the unit operator; the second diagram is relevant to
the contribution of quark condensate, obviously one can neglect the
heavy-quark condensate at all. The Wilson coefficient of the
two-gluon condensate operator is obtained from Fig.~\ref{wilson
coefficients graph}(c-h). The last two diagrams in Fig.~\ref{wilson
coefficients graph}(i-j) stand for the contribution of quark-gluon
mixing condensate. In this work, all of the Wilson coefficients are
calculated up to the lowest order in the running coupling constant
$\alpha_s$ of strong interaction.

\begin{figure}[tb]
\begin{center}
\begin{tabular}{ccc}
\includegraphics[scale=0.8]{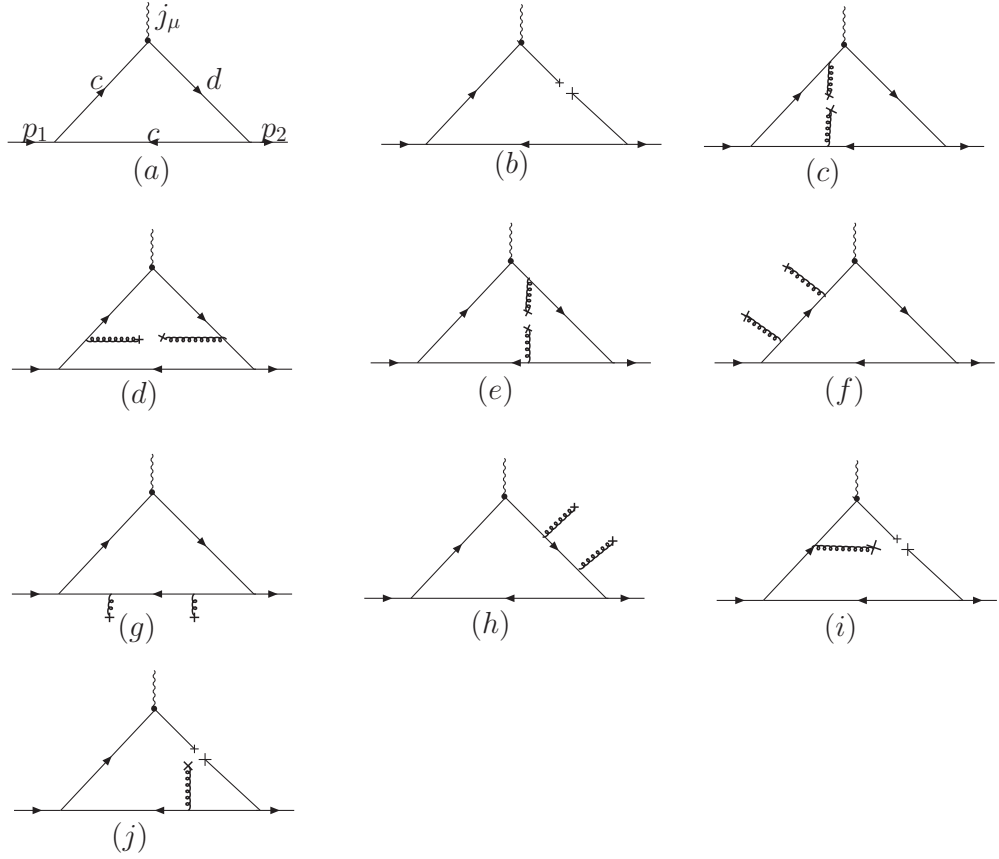}
\vspace{-6 cm}
\end{tabular}
\caption{Graphs for the Wilson coefficients in the operator product
expansion of the correlation function. (a) is for the contribution
of unit operator; (b) for the two-quark condensate; (c-h) describe
the contributions from gluon condensate, (i-j) is for the
quark-gluon mixing condensate.} \label{wilson coefficients graph}
\end{center}
\end{figure}

\subsubsection{The calculations of perturbative contributions to $\tilde{\Pi}_{\mu \nu}$}
The contribution of perturbative part to the three-point correlation
function $\tilde{\Pi}_{\mu \nu}$ comes from Fig. \ref{wilson
coefficients graph} (a), which can be expressed as
\begin{eqnarray}
\tilde{C}^{pert}_{\mu \nu}=i^2 \int {d^4 k \over (2 \pi)^4} (-1)
{\rm{Tr}}[\gamma_{\nu}{i \over \not k -m_c } i \gamma_5 {i \over
\not {p_2} + \not k -m_q} \sigma_{\mu \nu'}q^{\nu'}(1+\gamma_5) {i
\over \not {p_1} +\not k -m_c}]. \label{perturbative contribution 3}
\end{eqnarray}
Again, we need to express $\tilde{C}^{pert}_{\mu \nu}$ in the form
of dispersion integrals. Then, we arrive at the following expression
\begin{eqnarray}
\tilde{C}^{pert}_{\mu \nu}=\int\int ds_1 ds_2 {\tilde{\rho}^{pert}
_{\mu \nu} (s_1,s_2,q^2) \over (s_1-p_1^2) (s_2-p_2^2)}.
\end{eqnarray}
The integration region is determined by the following condition
\begin{eqnarray}
-1 \leq  {2 s_1 (s_2+m_c^2-m_q^2)-s_1 (s_1+s_2 -q^2) \over
\lambda^{1/2}(s_1,s_2,q^2) \lambda^{1/2}(m_c^2,s_1,m_c^2) }\leq 1,
\label{integral region}
\end{eqnarray}
where $\lambda(a,b,c)= a^2+b^2+c^2 -2ab-2ac-2bc $. Then, following
the standard approach,  putting all the internal quark lines on
their mass shells in terms of the Cutkosky's rules, we can derive
the spectral density $\tilde{\rho}^{pert}_{\mu \nu}$ as
\begin{eqnarray}
\tilde{\rho}^{pert}_{\mu \nu}=i{\tilde{\rho}}^{pert}_{0}
\epsilon_{\mu \nu \alpha \beta}
p_1^{\alpha}p_2^{\beta}+{\tilde{\rho}}^{pert}_1 {p_1}_{\mu}
{p_{1}}_{\nu} +{\tilde{\rho}}^{pert}_2 {p_2}_{\mu} {p_{2}}_{\nu}
+{\tilde{\rho}}^{pert}_3 {p_2}_{\mu}
{p_{1}}_{\nu}+{\tilde{\rho}}^{pert}_4 {p_1}_{\mu}
{p_{2}}_{\nu}+{\tilde{\rho}}^{pert}_5 g_{\mu \nu}, \label{spectral
density 3}
\end{eqnarray}
and the explicit expressions of ${\tilde{\rho}}^{pert}_{i}$ are
collected in  Appendix \ref{Wilson coefficients for D0} for the
concision of the text.

\subsubsection{The contribution of gluon condensate to $\tilde{\Pi}_{\mu \nu}$}
Now let us focus on the computation of the Wilson coefficient
corresponding to gluon condensate. In particular, it is worth to
emphasize that the contributions of gluon condensate to the
correlation function no longer vanish, even after performing the
double Borel transformation with respect to the variables $p_1^2$
and $p_2^2$. This point is an important difference between the sum
rules of vector current and tensor density. The calculations are
much the same as that for the case of ${\Pi}_{\mu \nu}$ in Ref.
\cite{Wang:2007ys}, and the only difference is that the weak decay
vertex ``$\gamma_{\mu}(1-\gamma_5)$" is replaced by the tensor one
``$\sigma_{\mu \nu}q^{\nu}(1+\gamma_5)$ ". Besides, we also need to
rewrite the Wilson coefficient in the form of dispersion integral as
that for the perturbative part, i.e.,
\begin{eqnarray}
\tilde{C}^{GG}_{\mu \nu}=\int\int ds_1 ds_2 {{\tilde{\rho}}^{GG}
_{\mu \nu} (s_1,s_2,q^2) \over (s_1-p_1^2) (s_2-p_2^2)},
\end{eqnarray}
with the integral region being the same as that for the perturbative
one.

The next step is to decompose the above spectral density
${\tilde{\rho}}^{GG} _{\mu \nu}$ into various Lorentz structures,
namely
\begin{eqnarray}
\tilde{\rho}^{GG}_{\mu \nu}=i{\tilde{\rho}}^{GG}_{0} \epsilon_{\mu
\nu \alpha \beta} p_1^{\alpha}p_2^{\beta}+{\tilde{\rho}}^{GG}_1
{p_1}_{\mu} {p_{1}}_{\nu} +{\tilde{\rho}}^{GG}_2 {p_2}_{\mu}
{p_{2}}_{\nu} +{\tilde{\rho}}^{GG}_3 {p_2}_{\mu}
{p_{1}}_{\nu}+{\tilde{\rho}}^{GG}_4 {p_1}_{\mu}
{p_{2}}_{\nu}+{\tilde{\rho}}^{GG}_5 g_{\mu \nu}, \label{spectral
density 3},
\end{eqnarray}
with the explicit expressions of ${\tilde{\rho}}^{GG}_{i}$ displayed
in Appendix \ref{Wilson coefficients for D0} for completeness of the
paper. The continuum subtraction should be carried out not only for
the perturbative diagram, but also for the contributions of the
gluon condensate.

\subsection{Wilson coefficients of the correlation function $\tilde{\Pi}_{\mu \nu \rho}$}
Now, we turn our attention to the operator product expansion for the
three-point function $\tilde{\Pi}_{\mu \nu \rho}$ in the deep
Euclidean region, which can be extended to the concerned physical
region analytically. Repeating the previous procedures but replacing
the vertex for a pseudoscalar meson to that for a vector meson, one
can immediately obtain the expressions of the Wilson coefficients
for all the concerned operators.
\subsubsection{Calculations of perturbative contributions to $\tilde{\Pi}_{\mu \nu \rho}$}
We can write the perturbative contribution to $\tilde{\Pi}_{\mu \nu
\rho}$ shown in Fig. \ref{wilson coefficients graph} (a) as
\begin{eqnarray}
C^{pert}_{\mu \nu \rho}=i^2 \int {d^4 k \over (2 \pi)^4} (-1)
{\rm{Tr}}[\gamma_{\nu}{i \over \not k -m_c } \gamma_{\rho} {i \over
\not {p_2} + \not k -m_q} \sigma_{\mu \nu'}q^{\nu'}(1+\gamma_5) {i
\over \not {p_1} +\not k -m_c}].
\end{eqnarray}
The perturbative part should be expressed in the form of dispersion
integral for performing an efficient subtraction of the continuum
states. In other words, we have
\begin{eqnarray}
\tilde{C}^{pert}_{\mu \nu \rho}=\int\int ds_1 ds_2
{{\tilde{\rho}}^{pert} _{\mu \nu \rho} (s_1,s_2,q^2) \over
(s_1-p_1^2) (s_2-p_2^2)},
\end{eqnarray}
where the integral region is the same as before. Following the
standard approach, then, we can analyze the spectral function for
the perturbative part as below
\begin{eqnarray}
\tilde{\rho}^{pert}_{\mu \nu \rho} &=& {{\tilde{\rho'}}}^{pert}_{1}
\epsilon_{\mu \nu \rho \lambda}p_1^{\lambda}+
{{\tilde{\rho'}}}^{pert}_{4} \epsilon_{\mu \nu \rho
\lambda}p_2^{\lambda}+ {{\tilde{\rho'}}}^{pert}_{5} \epsilon_{\mu
\nu \alpha \beta} p_1^{\alpha} p_1^{\beta}
{p_1}_{\nu}+i{{\tilde{\rho'}}}^{pert}_{7}g_{\mu \nu}{p_1}_{\rho}
+i{{\tilde{\rho'}}}^{pert}_{9}g_{\nu \rho} {p_1}_{\mu} \nonumber \\
&&+i{{\tilde{\rho'}}}^{pert}_{11}g_{\mu \rho}
{p_2}_{\nu}+i{{\tilde{\rho'}}}^{pert}_{15}g_{\nu
\rho}{p_2}_{\mu}+...,
\end{eqnarray}
where only the structures related to the form factors are listed for
simplification. Furthermore, the explicit forms of
${{\tilde{\rho'}}}^{pert}_{i}$ which are tedious, can be found in
Appendix \ref{Wilson coefficients for D0 star}.

\subsubsection{The calculation of gluon condensate to $\tilde{\Pi}_{\mu \nu \rho}$}
Now we  concentrate on the calculations of the Wilson coefficient of
gluon condensate for $\tilde{\Pi}_{\mu \nu \rho}$. The Wilson
coefficient is not equal to zero for the gluon condensate in the
operator expansion of $\tilde{\Pi}_{\mu \nu \rho}$. The dispersion
integral for this Wilson coefficient can be written as
\begin{eqnarray}
\tilde{C}^{GG}_{\mu \nu \rho}=\int\int ds_1 ds_2
{{\tilde{\rho}}^{GG} _{\mu \nu \rho} (s_1,s_2,q^2) \over (s_1-p_1^2)
(s_2-p_2^2)},
\end{eqnarray}
with the integral region being the same as that for the perturbative
part.

Next, we can decompose the above spectral density into various
Lorentz structures as
\begin{eqnarray}
\tilde{\rho}^{GG}_{\mu \nu \rho} &=& {{\tilde{\rho'}}}^{GG}_{1}
\epsilon_{\mu \nu \rho \lambda}p_1^{\lambda}+
{{\tilde{\rho'}}}^{GG}_{4} \epsilon_{\mu \nu \rho
\lambda}p_2^{\lambda}+ {{\tilde{\rho'}}}^{GG}_{5} \epsilon_{\mu \nu
\alpha \beta} p_1^{\alpha} p_1^{\beta}
{p_1}_{\nu}+i{{\tilde{\rho'}}}^{GG}_{7}g_{\mu \nu}{p_1}_{\rho}
+i{{\tilde{\rho'}}}^{GG}_{9}g_{\nu \rho} {p_1}_{\mu} \nonumber \\
&& +i{{\tilde{\rho'}}}^{GG}_{11}g_{\mu \rho}
{p_2}_{\nu}+i{{\tilde{\rho'}}}^{GG}_{15}g_{\nu \rho}{p_2}_{\mu}+...,
\end{eqnarray}
where the explicit forms of ${{\tilde{\rho'}}}^{GG}_{i}$ are given
in Appendix \ref{Wilson coefficients for D0 star}.

\section{Numerical analysis of form factors in QCD sum rules}
\label{Numerical results}

Eventually we are able to calculate the form factors numerically.
Firstly, we explicitly present all the input parameters which are
adopted in our numerical computations, as below \cite{ioffe 2, PDG,
korner}
\begin{equation}
\begin{array}{ll}
m_c(m_c)=1.275 \pm 0.015 \rm{GeV}, & m_u(1 {\rm{GeV}})=2.8
{\rm{MeV}},
\\
\alpha_{s}(1 {\rm{GeV}})=0.517, & m_{J/\psi}=3.097 \rm{GeV},
\\
 m_{D^{0}}=1.865 \rm{GeV}, &   m_{D^{*0}}=2.007 \rm{GeV},
\\
f_{J/\psi}=337 ^{+12}_{-13} \rm{MeV},    & f_{D^0}=166^{+9}_{-10}
\rm{MeV},
\\
f_{D^{0*}}=240^{+10}_{-10}  \rm{MeV},  & \langle {\alpha_s \over
\pi} G_{\mu \nu}^a G^{a \mu \nu}\rangle=(0.005 \pm
0.004){\rm{GeV}}^4,
\\
\langle \bar{u} u\rangle \cong -(1.65 \pm 0.15) \times
10^{-2}{\rm{GeV}}^3.
\end{array}
\end{equation}

All the QCD parameters are set at the renormalization scale around 1
GeV.   To reduce theoretical uncertainties in the three-point sum
rules of the weak transition form factors, due to  masses of quarks,
threshold parameters and Coulomb-like corrections for $J/\psi$
effectively \cite{kiselev Bc},  we apply the decay constants
$f_{J/\psi}$ and $f_{\bar{D}^{(*)0}}$ which are calculated with the
two-point QCD sum rules up to the leading order of $\alpha_s$, to
the three-point sum rules.  The details about the  calculations of
the decay constants of both $J/\psi$ and $\bar{D}^{(*)0}$ in the
framework of QCD sum rules,   are presented in
Ref.\cite{Wang:2007ys}.

For determining the threshold parameters $s_1^0$ and $s_2^0$, one
demands the QCD sum rules results to be sufficiently stable with
respect to variation of $M_1^2$ and $M_2^2$ within relatively large
regions, and their values  should be around the mass squares of the
corresponding first excited states. As for the heavy-light mesons,
the standard value of the threshold in an $X$ channel should be
$s^0_{X}=(m_X+\Delta_X)^2$, where $\Delta_X$ is about $0.6$ GeV
\cite{dosch, matheus, bracco, navarra, Colangelo}, and we simply
take it as $(0.6 \pm 0.1)\; \mathrm{GeV}$ for the error estimate in
our numerical analysis. For the heavy charmonium, following the
method given in Ref. \cite{matheus, bracco, Colangelo}, we select an
effective threshold parameter to ensure the appearance of a
satisfactory platform which is around the mass square of $\psi(2S)$.
In this way, the contributions from both the excited states
including $\psi(2S)$ and the continuum states are contained in the
spectral function.

\subsection{Analysis on the sum rules for the form factors}

\subsubsection{Evaluation of the sum rules for the $J/\psi \to \bar{D}^{0}$}

With   all the parameters listed above, we can obtain the numerical
values of the form factors. The form factors should not depend on
the Borel masses $M_1$ and $M_2$ in a complete theory. However, as
we truncate the operator product expansion up to dimension-5 and
keep the perturbative expansion in $\alpha_s$ to the leading order,
dependence of the form factors on these two Borel parameters would
emerge. Therefore, one should look for a region(s) where the results
only mildly vary with respect to the Borel masses, so that the
truncation is reasonable and acceptable.

With a careful analysis, $s_1^0=13.7$ GeV$^2$ and $s_2^0=6.1$
GeV$^2$  are chosen  for the sum rules of  form factors $T_i$
($i=1,2,3$).  As commonly understood, the Borel parameters $M_1^2$
and $M_2^2$ should not be too large in order to ensure that the
contributions from the higher excited states and continuum are not
too significant. On the other hand, the Borel masses also could not
be too small for the sake of validity of OPE in the deep Euclidean
region, since the contributions of higher dimension operators
pertain to the higher orders in ${1 \over M_i}(i=1,2) $. Unlike the
treatment adopted in previous literature \cite{p. ball,weak decays
of QCDSR 1} where the ratio of $M_1$ and $M_2$ was fixed, in this
paper, when calculating the form factors, we let $M_1$ and $M_2$
vary independently as suggested by the authors of Ref.
\cite{yangkc,kiselev Bc}.

\begin{figure}[tb]
\begin{center}
\begin{tabular}{ccc}
\includegraphics[scale=0.5]{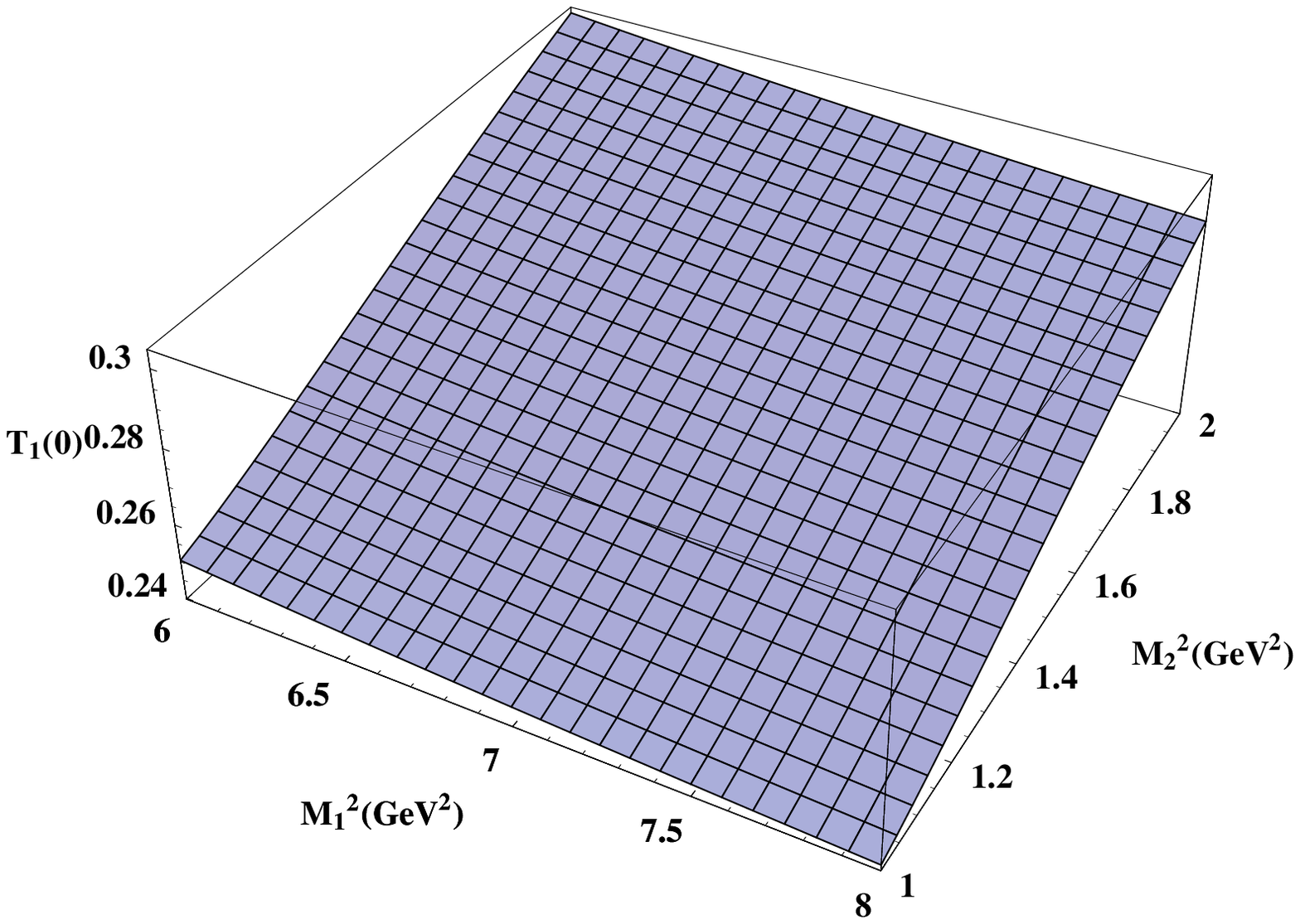}
\includegraphics[scale=0.5]{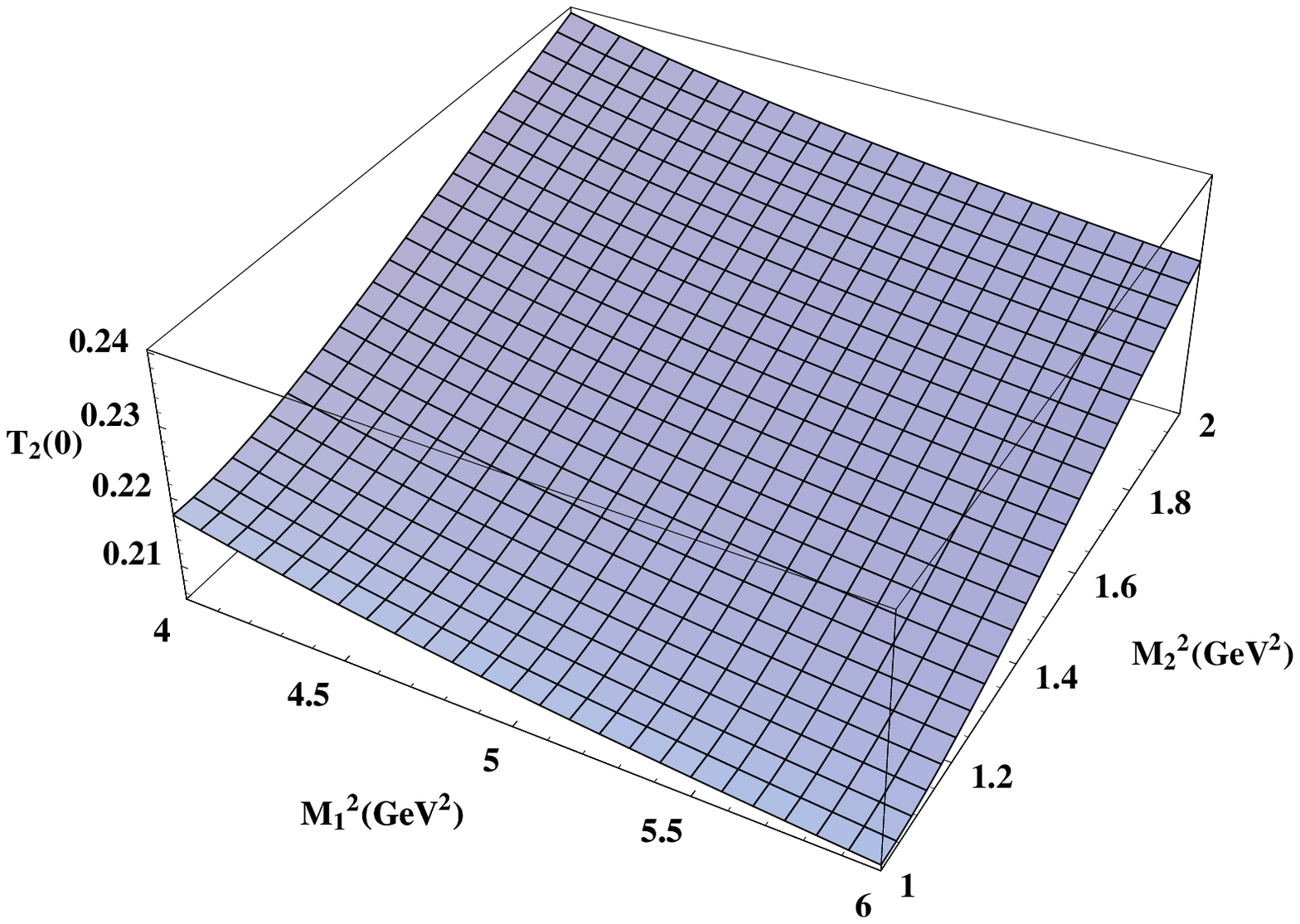}
\\
\includegraphics[scale=0.5]{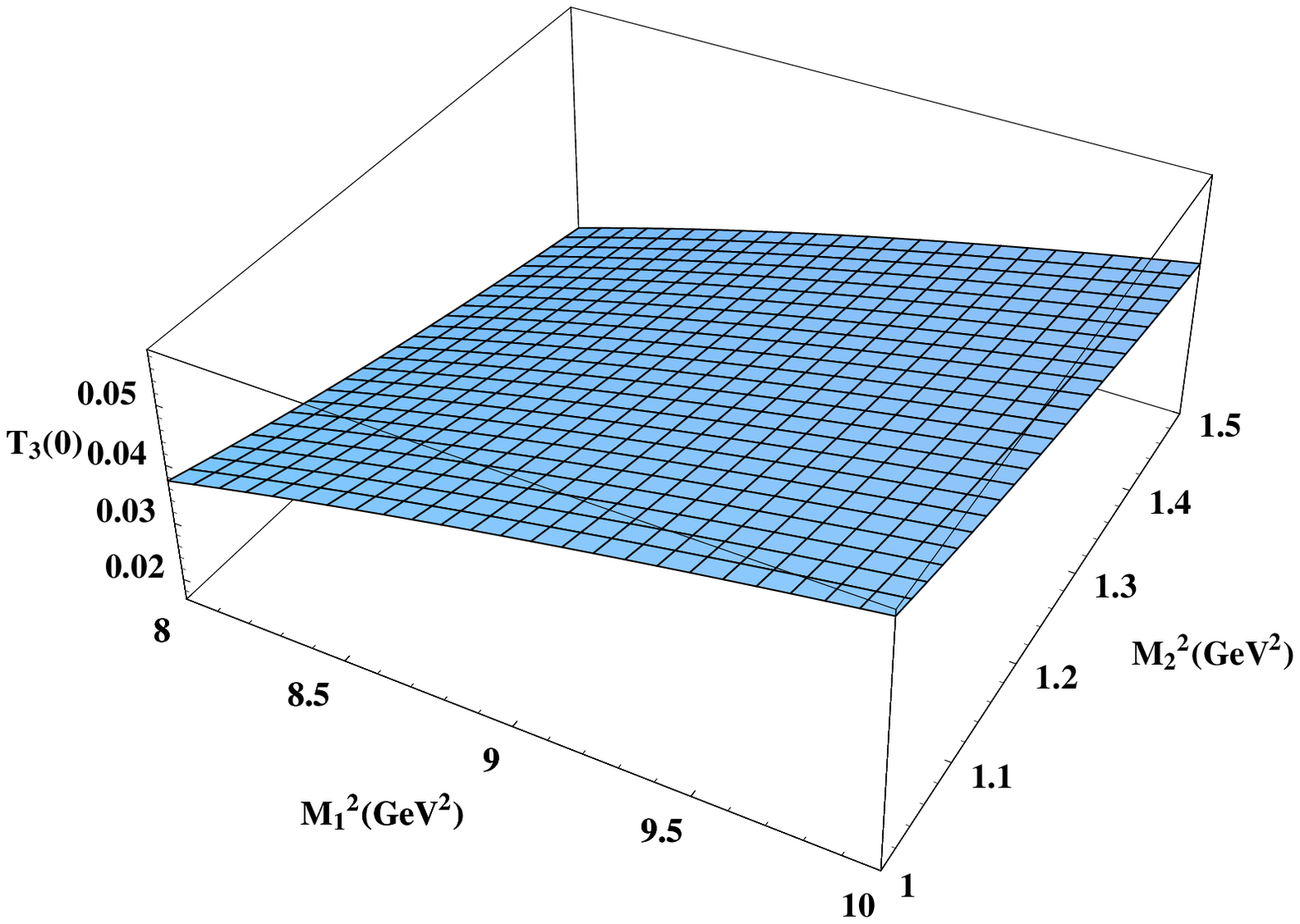}
\end{tabular}
\caption{various form factors $T_1, T_2$ and $T_3$ at $q^2=0$
responsible for the decay of $J/\psi \to \bar{D}^{0} $within the
Borel window. }\label{all form factor of J psi to D0}
\end{center}
\end{figure}

As observed in last section, the contributions of gluon condensate
are nontrivial for tensor density, that is different from the case
of vector current. We display the form factors at zero momentum
transfer in Fig. \ref{all form factor of J psi to D0}. As for the
form factor $T_1$, the Borel masses are set as $M_1^2 \in [6.0, 8.0]
\mathrm{GeV}^2, M_2^2 \in [1.0, 2.0] \mathrm{GeV}^2$, according to
the condition that contributions from both the continuum states and
the non-perturbative gluon condensate to the total sum rules are no
more than 40 \% and then $T_1(q^2=0)=0.27^{+0.04}_{-0.04}$ is
resulted in. Here we have combined the errors induced by the
variations of Borel masses, threshold values, the mass of charm
quark,  decay constants of involved mesons as well as gluon
condensate. The value of $T_2$ is set as $0.22^{+0.02}_{-0.03}$ by
the constraint that contributions of neither the higher states nor
the dimension-4 gluon condensate can exceed  40 \% of the total
contribution to the whole sum rules, and it determines the Borel
region as $M_1^2 \in [4.0, 6.0] \mathrm{GeV}^2, M_2^2 \in [1.0, 2.0]
\mathrm{GeV}^2$. Finally, the Borel masses for the form factor $T_3$
are chosen as $M_1^2 \in [8.0, 10.0] \mathrm{GeV}^2, M_2^2 \in [1.0,
1.5] \mathrm{GeV}^2$ under the requirement that the contributions
from both the continuum states and the gluon condensate should be
less than 40 \%, thus we have $T_3(q^2=0)$ as
$0.037^{+0.033}_{-0.037}$.

Naturally, we continue to investigate $q^2$ dependence of the form
factors at the region $q^2 \in [0, 0.47] \rm{GeV}^2$ evading the
non-Landau-type singularities. The results are shown in Fig. \ref{t
dependence of J psi to D0} within the given Borel region. We fit the
form factors with the double-pole approximation for phenomenological
applications. Here, one notices that the form factor $T_3$ decreases
quickly with the increase of the momentum transfer as $q^2
> 0.3 \rm{GeV}^2$, that is a consequence of the naive and artifact treatment of
the continuum density in our model and also owing to the smaller gap
for the kinematical threshold $(q^2)_{max}= 1.5 \rm{GeV}^2$ which
may spoil the operator product expansion \cite{weak decays of QCDSR
1}. Besides, we can also find that the $q^2$ dependence of $T_3$ at
the region $q^2 \in [0, 0.2] \rm{GeV}^2$  is rather mild with the
changes of the momentum transfer due to a cancelation between the
increase of the perturbative part and the decrease of the gluon
condensate, the dependence is shown in Fig. \ref{cancelation of
perturbative and GG T3 D0}. The other two form factors $T_1$ and
$T_2$ can also be written in  the double-pole form, namely
\begin{eqnarray}
F_{i}(q^2)={F_i(0) \over 1-a_i q^2/m_{\bar{D}^{0}}^{2}+b_i
q^4/m_{\bar{D}^{0}}^{4}},
\end{eqnarray}
where the parameters $a_i$ and $b_i$ can be determined from the
results given by the QCD sum rules in the region $q^2 \in [0, 0.47]
\rm{GeV}^2$ as
\begin{eqnarray}
a_{T_1}=1.70^{+0.23}_{-0.10}, \qquad b_{T_1}=0.44^{+0.56}_{-0.22},  \nonumber \\
a_{T_2}=0.75^{+0.35}_{-0.18}, \qquad b_{T_2}=0.40^{+0.31}_{-0.13},
\end{eqnarray}
with
\begin{eqnarray}
T_1(0)&=&0.27^{+0.04}_{-0.04}, \,\,\,T_2(0)=0.22^{+0.02}_{-0.03}.
\end{eqnarray}

\begin{figure}[tb]
\begin{center}
\begin{tabular}{ccc}
\includegraphics[scale=0.65]{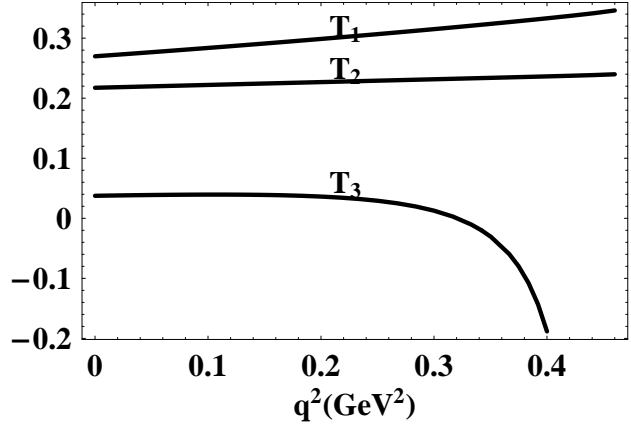}
\end{tabular}
\caption{$q^2$ dependence of form factors $T_1$, $T_2$ and $T_3$ for
the decay of $J/\psi \to \bar{D}^{0}$ within region without
non-Landau-type singularities.}\label{t dependence of J psi to D0}
\end{center}
\end{figure}

\begin{figure}[tb]
\begin{center}
\begin{tabular}{ccc}
\includegraphics[scale=0.65]{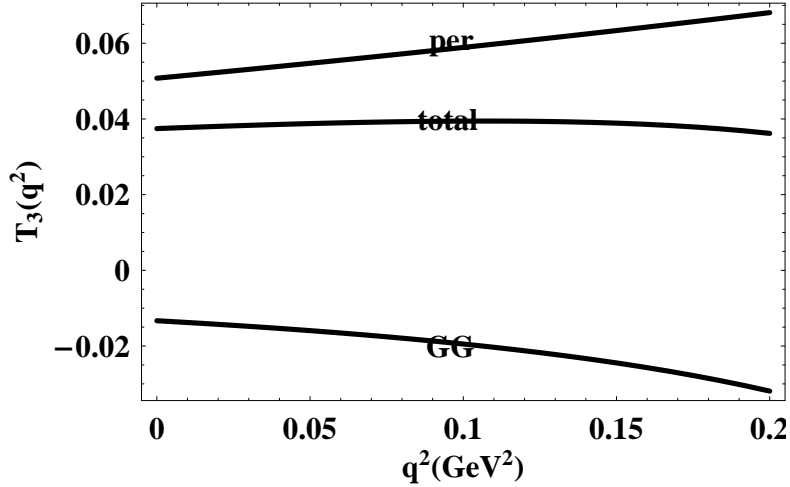}
\end{tabular}
\caption{mutual cancelation of $q^2$ dependence of form factor $T_3$
associating with the $J/\psi \to \bar{D}^{0}$ decay mode from the
perturbative part and gluon condensate.}\label{cancelation of
perturbative and GG T3 D0}
\end{center}
\end{figure}

\subsubsection{Estimation of the sum rules for the $J/\psi \to \bar{D}^{*0}$}

Now we numerically evaluate the sum rules for the $J/\psi \to
\bar{D}^{*0}$ based on the standard method. Obviously, the hadronic
matrix element for $J/\psi \to \bar{D}^{*0}$ is equal to that for
$J/\psi \to \bar{D}^{*-}$, as long as isospin violation effects can
be neglected. The threshold used here for the $\bar{D}^{*0}$ is also
the same as that for the $\bar{D}^{*-}$, i.e. $s_2^0=6.8
\rm{GeV}^2$. To start with, we study  form factors at zero momentum
transfer with the Borel masses presented in Fig. (\ref{all form
factor of J psi to D0 star A}-\ref{all form factor of J psi to D0
star V}). For the form factor $\tilde{T}_1$, the Borel platform is
taken as $M_1^2 \in [6.0, 10.0] \mathrm{GeV}^2, M_2^2 \in [1.0, 1.6]
\mathrm{GeV}^2$ in agreement with the condition that the
contributions from both the continuum states and the gluon
condensate should be less than 35 \%  of the total contribution, and
we obtain  $\tilde{T}_1(q^2=0)$ as $0.42^{+0.02}_{-0.03}$.
Similarly, we have $\tilde{T}_2(q^2=0)$ as $0.70^{+0.07}_{-0.09}$
and $\tilde{T}_3$ to be $1.02^{+0.17}_{-0.19}$ with the Borel masses
being $M_1^2 \in [6.5, 8.0] \mathrm{GeV}^2, M_2^2 \in [1.5, 2.0]
\mathrm{GeV}^2$. In addition, the form factor $\tilde{T}_4$ is
$0.20^{+0.01}_{-0.13}$ with the Borel region $M_1^2 \in [6.0, 10.0]
\mathrm{GeV}^2, M_2^2 \in [1.5, 2.5] \mathrm{GeV}^2$. Then
$\tilde{T}_5(q^2=0)$ with the Borel mass as $0.41^{+0.03}_{-0.02}$
GeV$^2$ and $\tilde{T}_6(q^2=0)=0.38^{+0.03}_{-0.02}$ with $M_1^2
\in [6.0, 10.0] \mathrm{GeV}^2, M_2^2 \in [1.0, 2.0]
\mathrm{GeV}^2$. Ultimately, we derive $\tilde{T}_7(q^2=0)=
0.11^{+0.01}_{-0.01}$ with the Borel platform $M_1^2 \in [6.0, 10.0]
\mathrm{GeV}^2, M_2^2 \in [1.5, 2.5] \mathrm{GeV}^2$.

With the form factors at zero momentum transfer, we can have their
values for non-zero $q^2$. We plot the form factors in the
kinematical region $q^2 \in [0, 0.42] \rm{GeV}^2$ free of
non-Landau-type singularities in Fig. \ref{t dependence of J psi to
D0 star}. It can be found that  $\tilde{T}_3$ increases quickly as
$q^2>0.3 \rm{GeV}^2$, while $\tilde{T}_5$ rises drastically as
$q^2>0.2 \rm{GeV}^2$. This point is similar to the behavior of $T_3$
responsible for  $J/\psi \to \bar{D}^{0}$, the reason was explained
in much detail there. The form factors $\tilde{T}_4$ and
$\tilde{T}_7$ can be fitted in the single-pole approximation
\begin{eqnarray}
 F_{i}(q^2)={F_i(0) \over (1-a_i
q^2/m_{\bar{D}^{*0}}^{2})},
\end{eqnarray}
while $\tilde{T}_1$, $\tilde{T}_3$ and $\tilde{T}_5$ can be written
in the following expression
\begin{eqnarray}
 G_{i}(q^2)={G_i(0) \over (1-a_i
q^2/m_{\bar{D}^{*0}}^{2})^2},
\end{eqnarray}
moreover,  $\tilde{T}_6$ is parameterized in the double-pole model
\begin{eqnarray}
H_{i}(q^2)={H_i(0) \over 1-a_i q^2/m_{\bar{D}^{*0}}^{2}+b_i
q^4/m_{\bar{D}^{*0}}^{4}}.
\end{eqnarray}
Similarly, the $q^2$ dependence of $\tilde{T}_2$ is extremely weak,
because the dominant contributions of perturbative part are almost
$q^2$ independent. The parameters $a_i$ and $b_i$ can be fixed in
terms of the results calculated with  QCD sum rules in the region
$q^2 \in [0, 0.42] \rm{GeV}^2$, then we can extend the above
expressions to the whole physical region $q^2 \in [0, 1.2]
\rm{GeV}^2$. The numbers of these parameters are given as
\begin{eqnarray}
a_{\tilde{T}_1}&=&1.70^{+0.13}_{-0.21}, \qquad a_{\tilde{T}_3}=2.32^{+0.06}_{-0.05},  \nonumber \\
a_{\tilde{T}_4}&=&0.84^{+0.31}_{-0.17}, \qquad a_{\tilde{T}_5}=2.76^{+0.28}_{-0.31}, \nonumber \\
a_{\tilde{T}_6}&=&1.95^{+0.33}_{-0.10}, \qquad b_{\tilde{T}_6}=2.14^{+0.34}_{-0.17}, \nonumber \\
a_{\tilde{T}_7}&=&2.00^{+0.21}_{-0.09},
\end{eqnarray}
and form factors at $q^2=0$ are summarized as
\begin{eqnarray}
\tilde{T}_1(0)&=&0.33^{+0.01}_{-0.01}, \,\,\,\tilde{T}_3(0)=0.80^{+0.09}_{-0.11}, \nonumber \\
\tilde{T}_4(0)&=&0.16^{+0.01}_{-0.01}, \,\,\,\tilde{T}_5(0)=0.32^{+0.02}_{-0.0}, \nonumber \\
\tilde{T}_6(0)&=&0.30^{+0.02}_{-0.01},
\,\,\,\tilde{T}_7(0)=0.089^{+0.007}_{-0.003}.
\end{eqnarray}

\begin{figure}[tb]
\begin{center}
\begin{tabular}{ccc}
\includegraphics[scale=0.50]{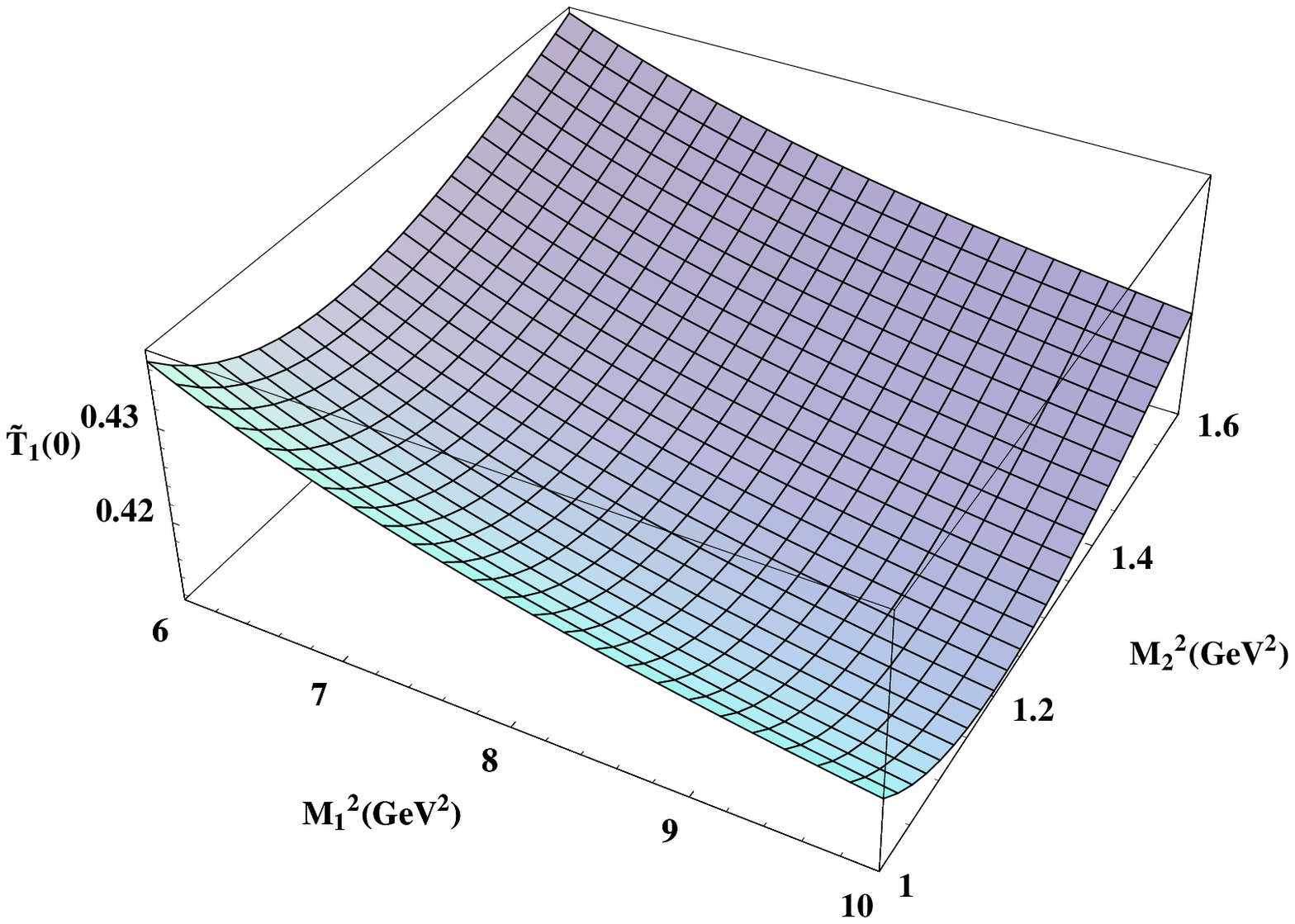}
\includegraphics[scale=0.50]{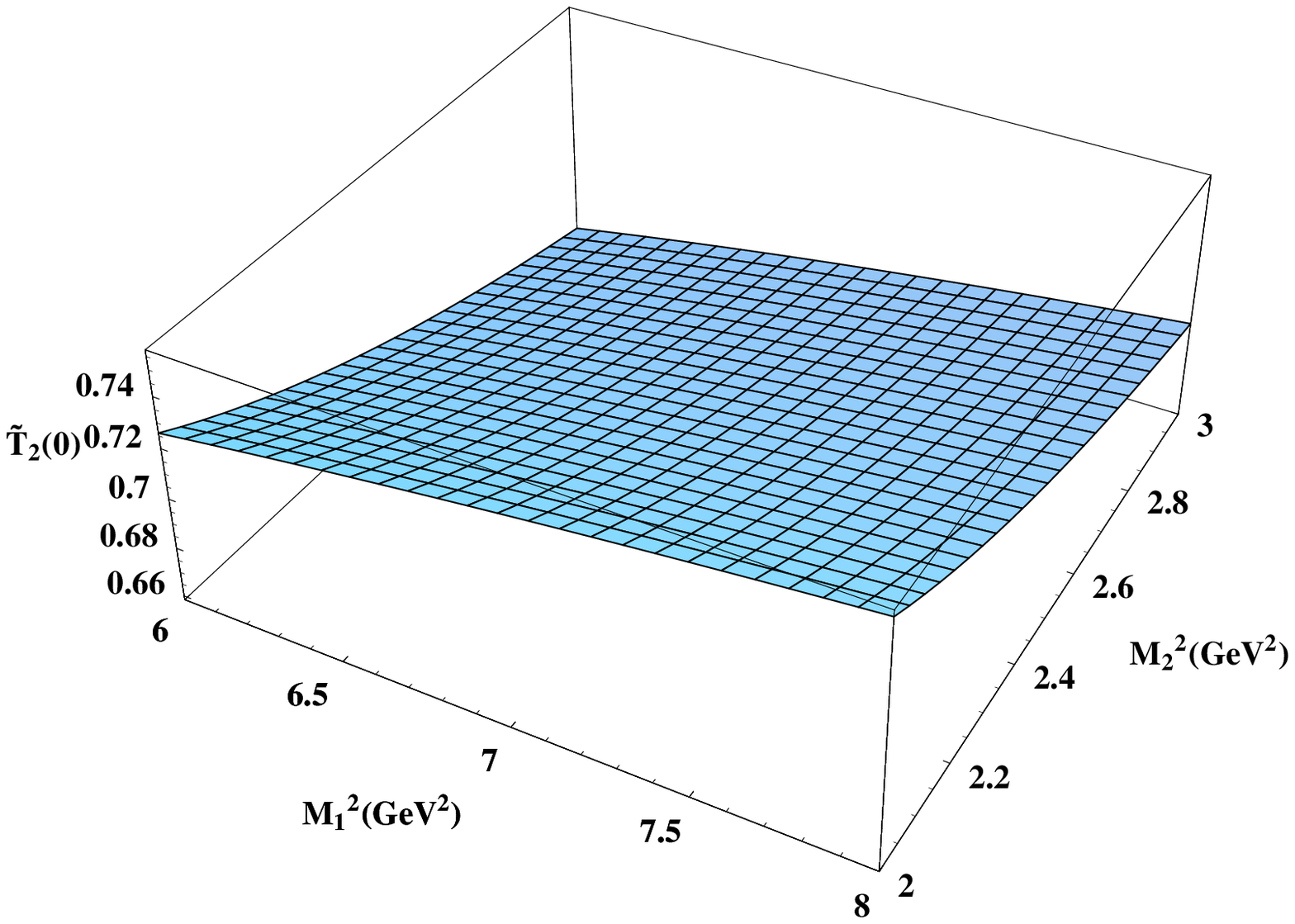}
\\
\includegraphics[scale=0.50]{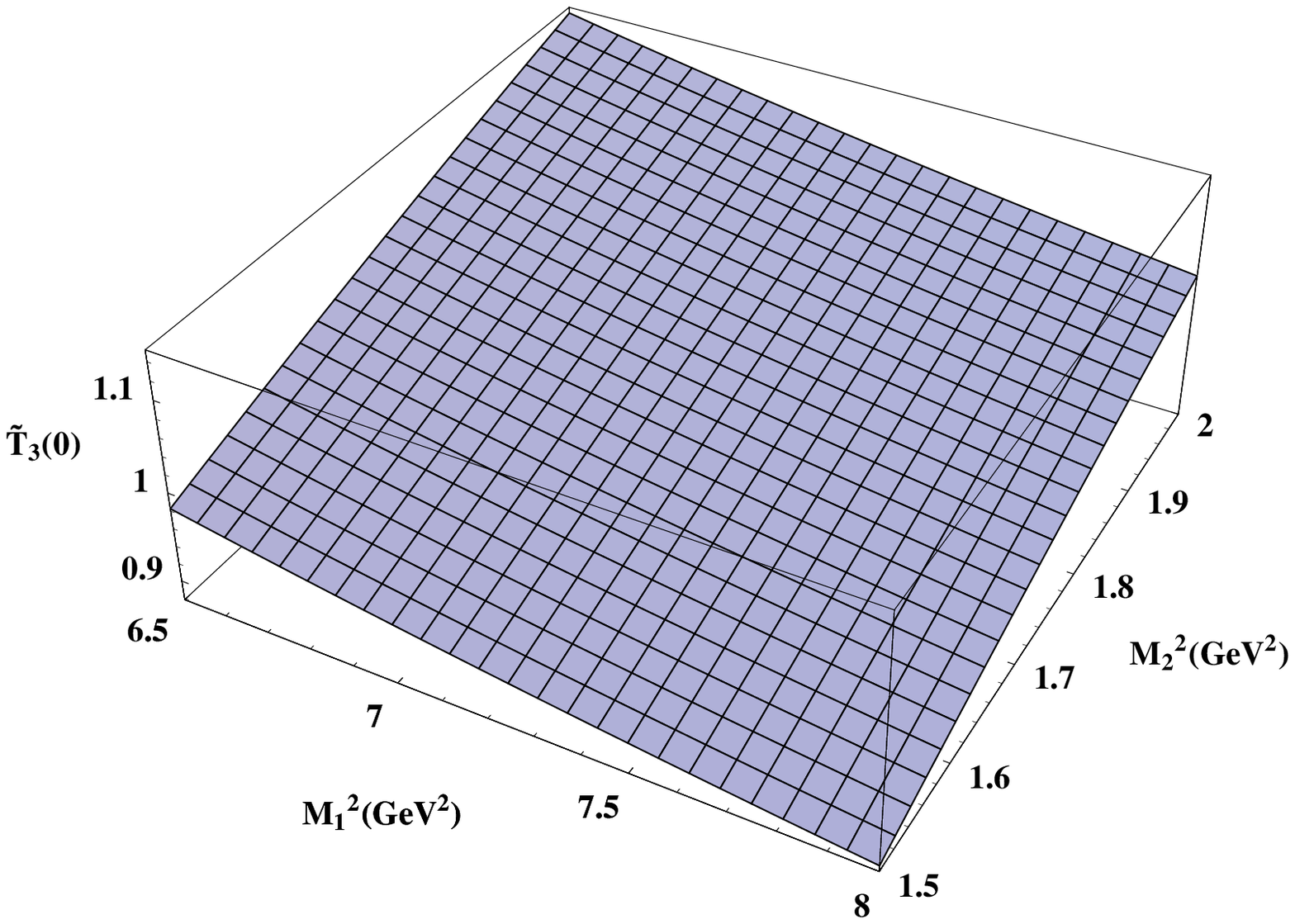}
\end{tabular}
\caption{various form factors $\tilde{T}_1$, $\tilde{T}_2$ and
$\tilde{T}_3$ at $q^2=0$ responsible for the decay of $J/\psi \to
\bar{D}^{*0} $within the Borel window. }\label{all form factor of J
psi to D0 star A}
\end{center}
\end{figure}

\begin{figure}[tb]
\begin{center}
\begin{tabular}{ccc}
\includegraphics[scale=0.50]{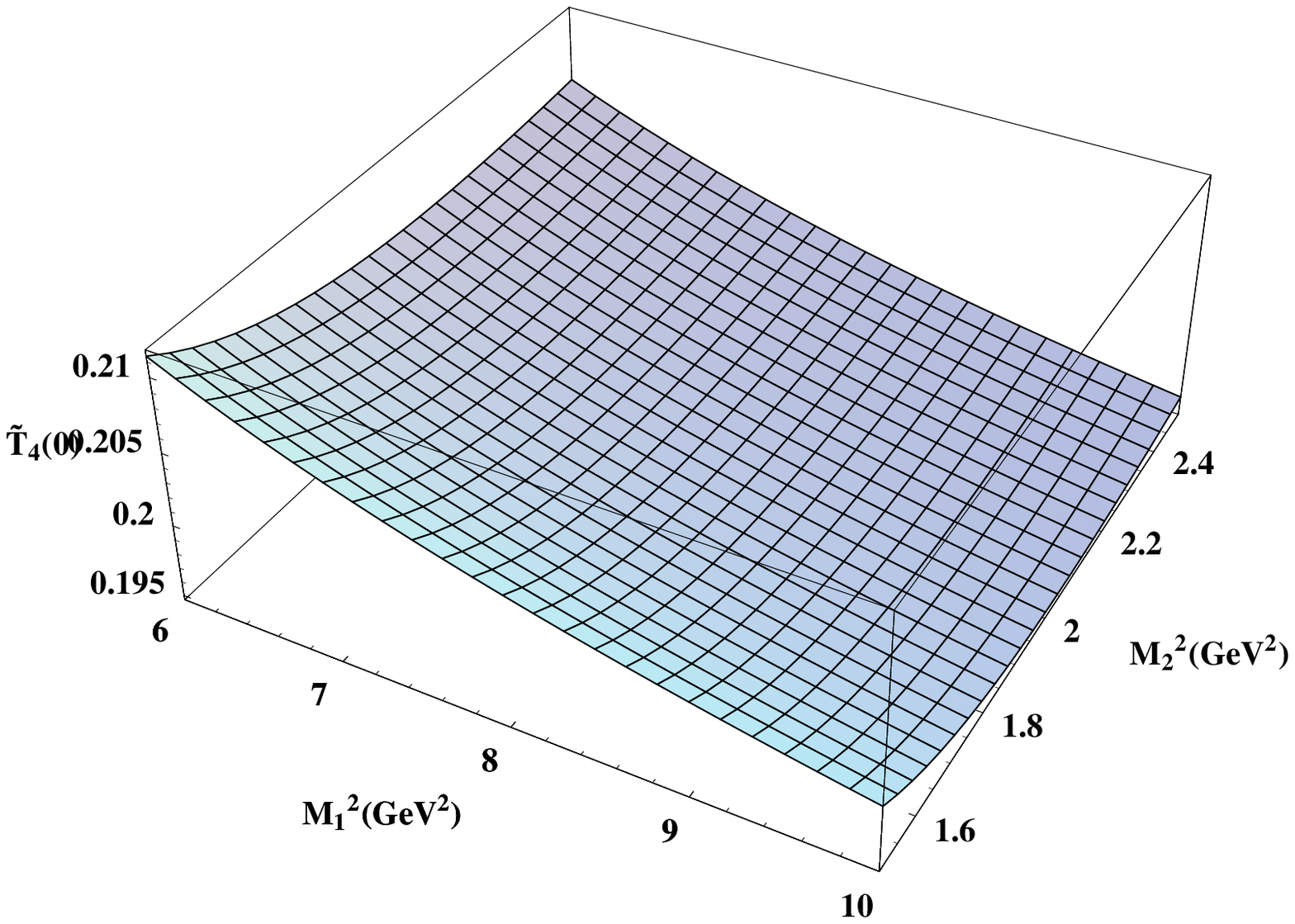}
\includegraphics[scale=0.50]{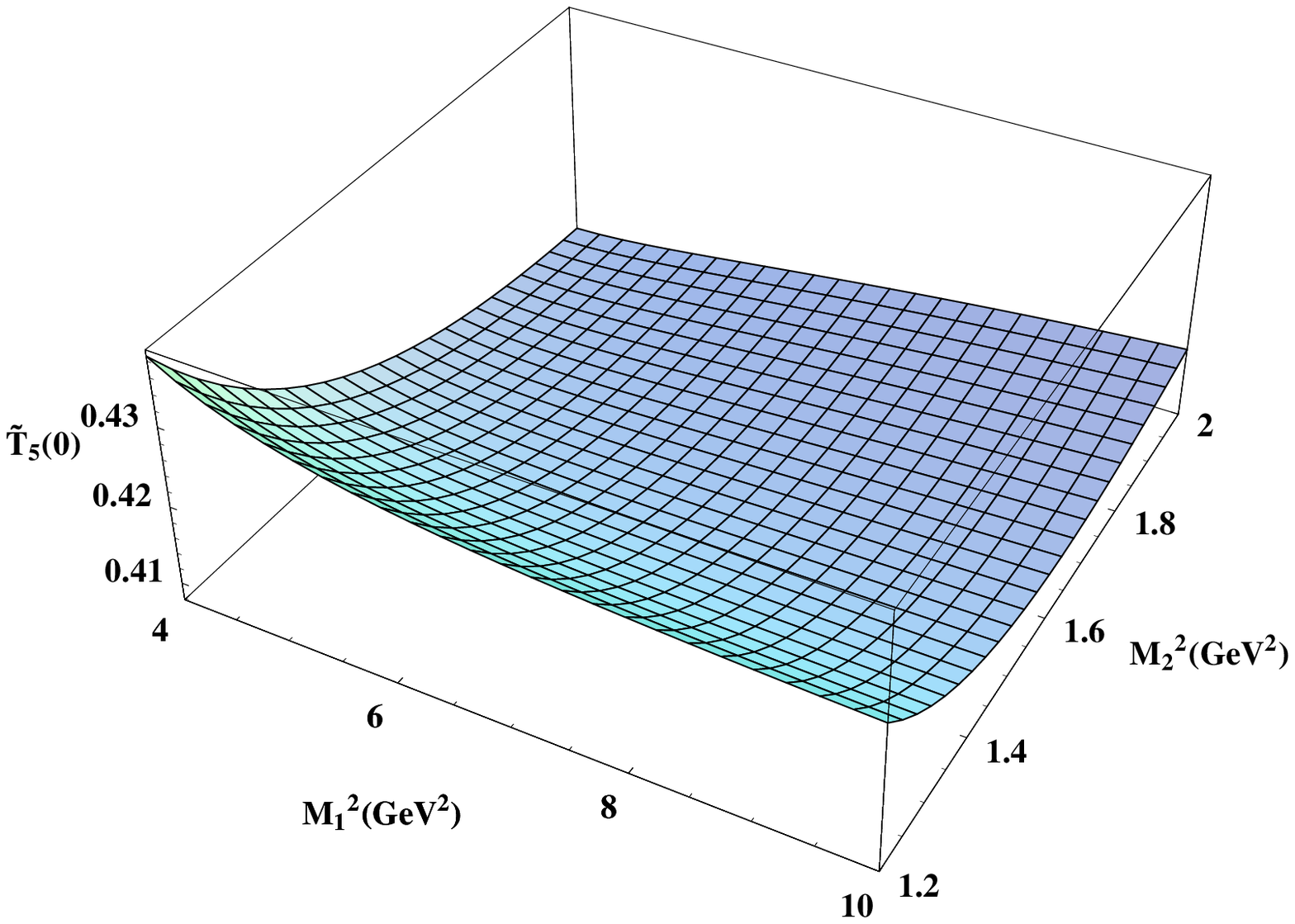}
\\
\includegraphics[scale=0.50]{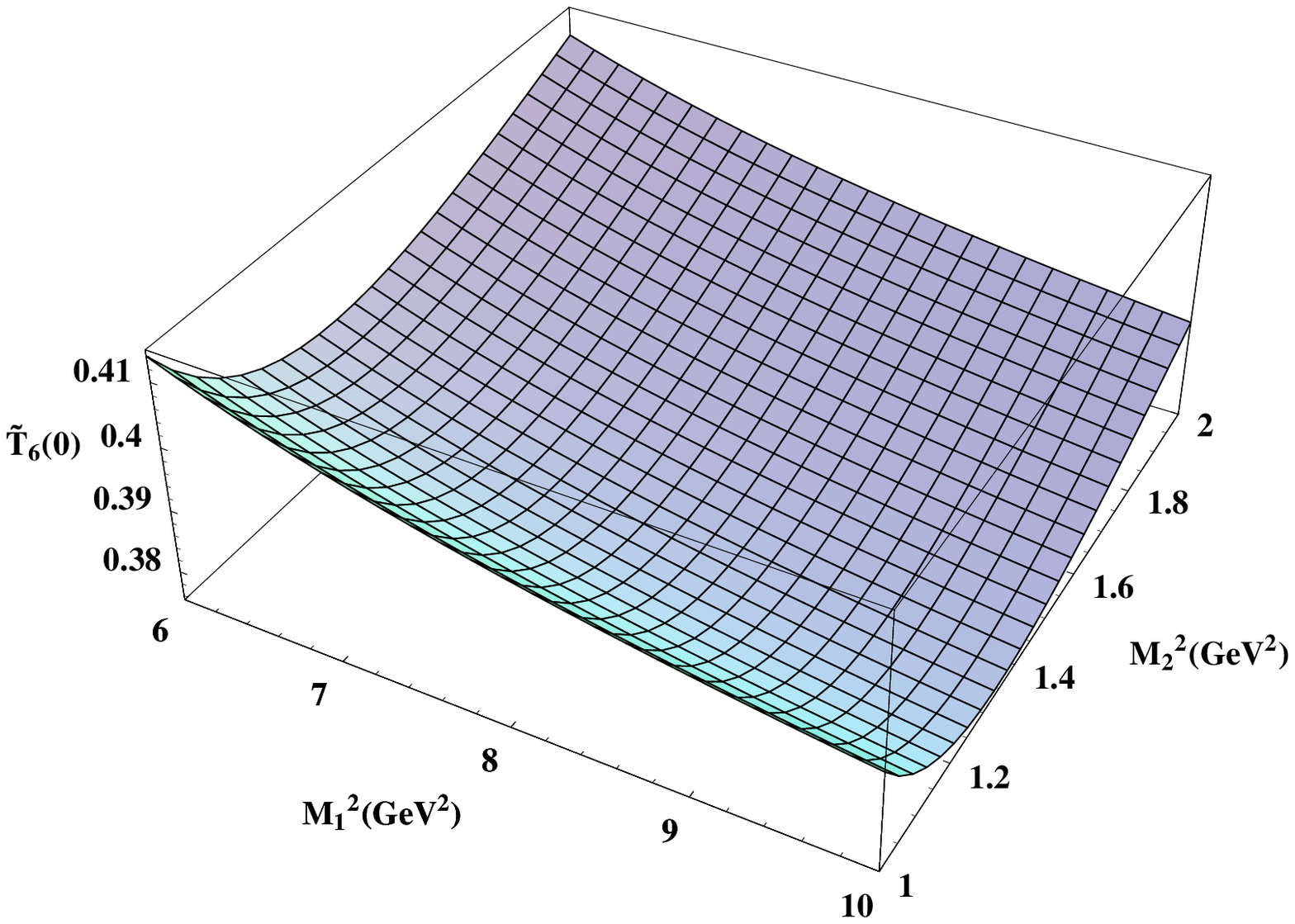}
\includegraphics[scale=0.50]{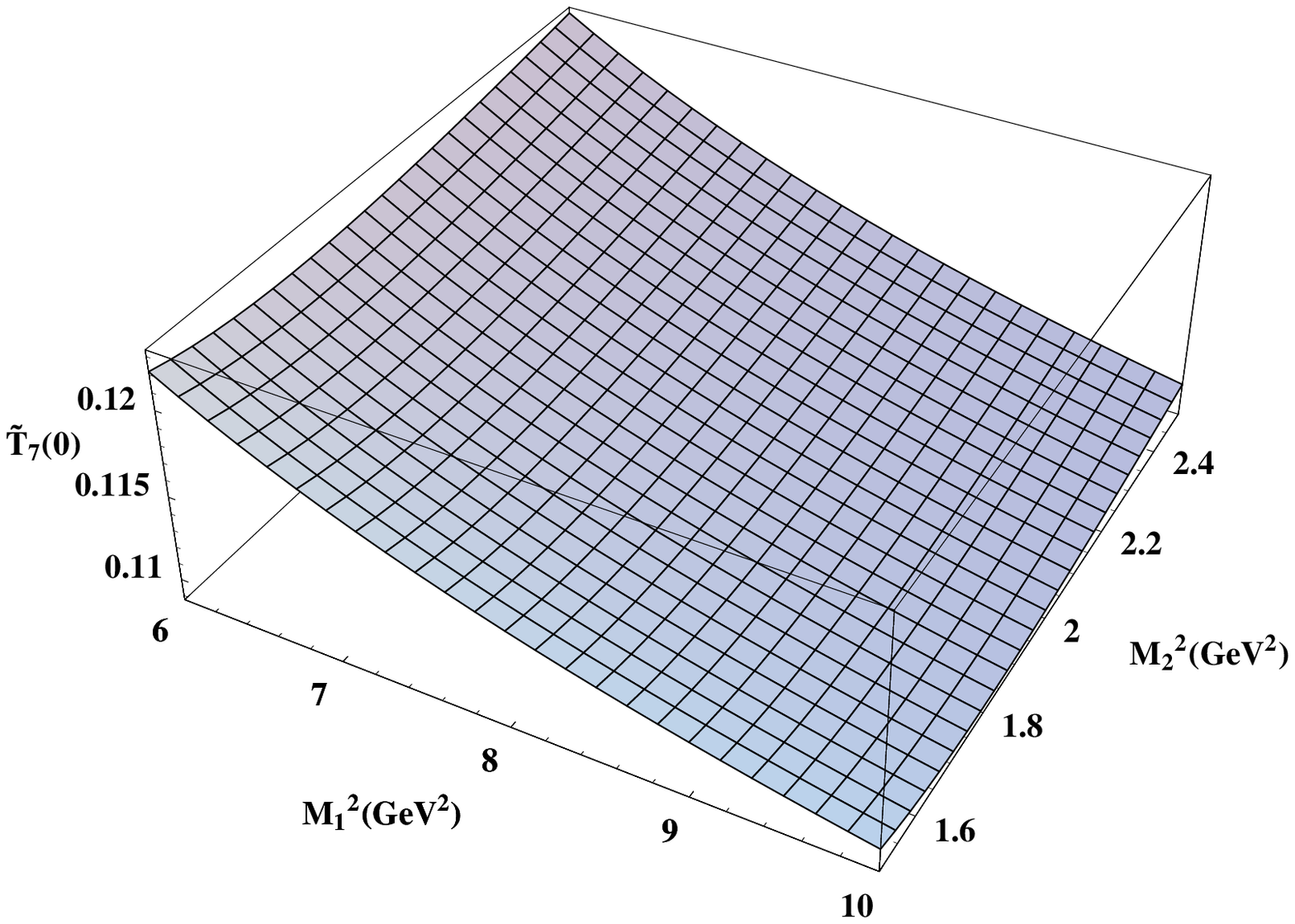}
\end{tabular}
\caption{various form factors $\tilde{T}_4$, $\bar{T}_5$,
$\tilde{T}_6$ and $\tilde{T}_7$ at $q^2=0$ responsible for the decay
of $J/\psi \to \bar{D}^{*0} $within the Borel window. }\label{all
form factor of J psi to D0 star V}
\end{center}
\end{figure}

\begin{figure}[tb]
\begin{center}
\begin{tabular}{ccc}
\includegraphics[scale=0.6]{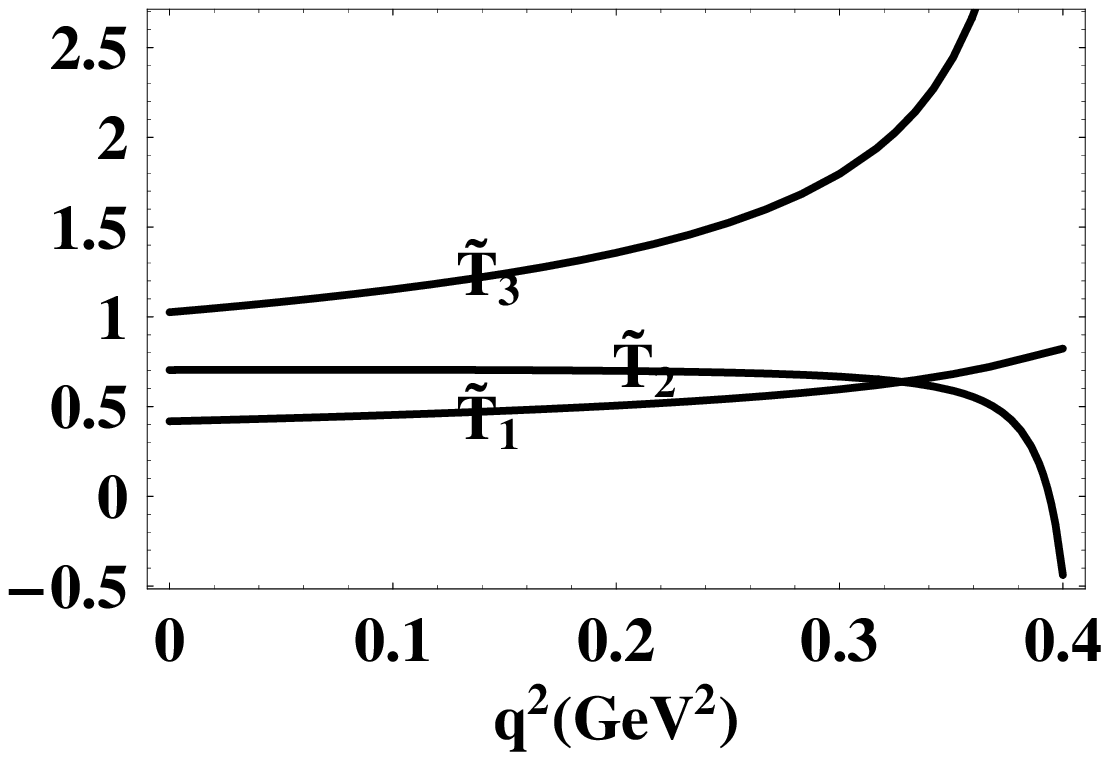}
\includegraphics[scale=0.6]{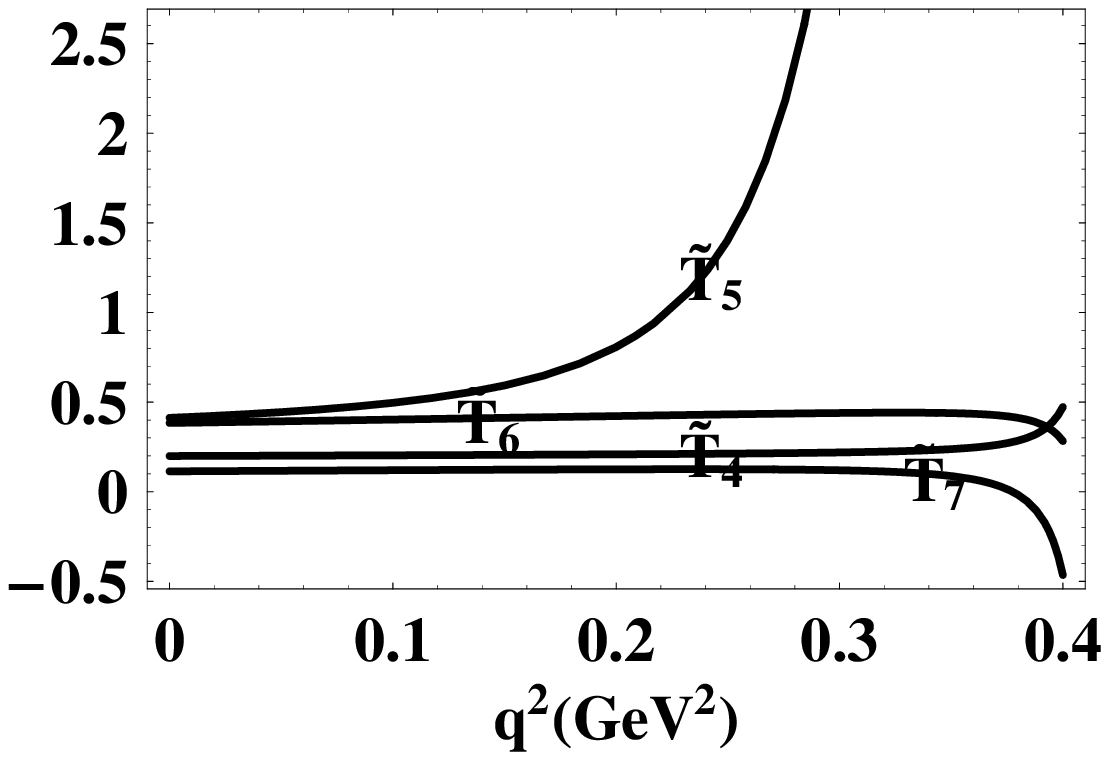}
\end{tabular}
\caption{$q^2$ dependence of form factors $\tilde{T}_1$,
$\tilde{T}_2$, $\tilde{T}_3$, $\tilde{T}_4$, $\tilde{T}_5$,
$\tilde{T}_6$ and $\tilde{T}_7$ for $J/\psi \to \bar{D}^{*0}$ within
the kinematical region without non-Landau-type
singularities.}\label{t dependence of J psi to D0 star}
\end{center}
\end{figure}

\subsection{Discussions on the theoretical uncertainties}
This subsection is devoted to a brief discussion about uncertainties
in our calculations.  One can observe that the errors originating
from neither the variations of the  Borel masses, nor the thresholds
for both $J/\psi$ and charmed meson channels exceed a level of $10
\%$. The results of form factors are in proportion to the inverse of
decay constants of the charmed mesons and $J/\psi$, which can be
easily observed from the definitions of correlation functions. The
uncertainties from decay constants of charmed mesons and $J/\psi$
are at the level of 10 \%, which can bring up about 10 \%
uncertainty to form factors at zero momentum transfer, but it would
not affect the values of parameters $a_i$ and $b_i$ which are used
to parameterize the $q^2$ dependence of the form factors. Besides,
the error bars corresponding to the changes of condensate parameters
are tiny, since only gluon condensate can contribute to the sum
rules, whose effect is at the order of a few percents for tensor
density transition. Moreover, one can also see  that the corrections
from the light quark masses are not significant because the role of
light quark masses is suppressed by a much larger energy scale of
Borel masses. In addition, the uncertainty caused by the charm quark
mass is at the one percent level \cite{ioffe 2}. We neglect the
$O(\alpha_s)$ corrections to the perturbative part, which is
expected to be quite small, and should not result in a drastic shift
to the final results.

In principle, the results of the form factors presented here can be
further improved by including the non-local quark condensate
\cite{p. ball} together with $O({1 \over m_c})$ power correction.
The former correction can result in a non-vanishing contribution of
the diagram (b) in Fig. \ref{wilson coefficients graph} even after
performing the double Borel transformation on two variables $p_1^2$
and $p_2^2$, however, it almost has no effect on the decay rates,
but only can moderate the $q^2$ dependence of the form factors. The
physical explanation of this effect is that quarks in the physical
vacuum may have a non-vanishing momentum \cite{p. ball}. The latter
correction will lead to the non-zero contributions of heavy-quark
condensate and heavy quark-gluon mixing condensate to the sum rules
for the form factors.

\section{Numerical analysis of decay rates for $J/\psi \to \bar{D}^{(*)0} l ^+ l^-$}
\label{decay rate}

With the form factors obtained above, we can calculate decay rates
of semi-leptonic decays $J/\psi \to D^{(*)-}_{d,s}+\bar l\nu$ and
$\bar{D}^{(*)0}\bar l l$. The related parameters are listed below
\cite{PDG,buras, p.ball:vector,Kuhn:2007vp}:
\begin{equation}
\begin{array}{ll}
G_{F}=1.166 \times 10^{-5}{\rm{GeV}}^{-2}, &
{\rm{sin}}^2\Theta_W=0.231,\nonumber\\
m_{W}=80.4 {\rm{GeV}}, &  m_b(m_b)=4.16 \pm 0.03 {\rm{GeV}}, \nonumber\\
m_e=0.51 \times 10^{-3}{\rm{GeV}}, & m_{\mu}=0.106 {\rm{GeV}}, \nonumber\\
|V_{ud}|=0.974, &  |V_{us}|=0.226 \pm 0.001,  \nonumber\\
|V_{ub}|=3.59 \pm 0.16 \times 10^{-3}, & |V_{cd}|=0.226 \pm 0.001, \nonumber\\
|V_{cs}|=0.973, &  |V_{cb}|=41.5^{+1.0}_{-1.1} \times 10^{-3}.
\end{array}
\end{equation}
For the semi-leptonic decay $J/\psi \to \bar{D}^{(*)0} l^{+}
l^{-}$($l= e, \mu$), the differential partial decay rate can be
written as
\begin{equation}
{d\Gamma_{J/\psi \to \bar{D}^{(*)0} l^{+} l^{-}} \over d q^2} ={1
\over 3}{1 \over (2 \pi)^3} {1 \over 32 m_{\psi}^3}
\int_{u_{min}}^{u_{max}} |{\widetilde{M}}_{J/\psi \to \bar{D}^{(*)0}
l^{+} l^{-}}|^2 du,
\end{equation}
where $u=(p_{\bar{D}^{(*)0}}+p_{l^+})^2$ and
$q^2=(p_{l^+}+p_{l^-})^2$; $p_{l^+}$ and $p_{l^-}$ are the momenta
of $l^{+}$ and $l^{-}$ respectively; the factor ``${1 \over 3}$''
comes from the average of the spin states of $J/\psi$;
$\widetilde{M}$ is the decay amplitude after integrating over the
angle between the $l^{+}$ and $\bar{D}^{(*)0}$.

The upper and lower bounds of $u$ are given as
\begin{eqnarray}
u_{max}&=&(E^{\ast}_{\bar{D}^{(*)0}}+E^{\ast}_{l^+})^2
-(\sqrt{E_{\bar{D}^{(*)0}}^{\ast 2}-m_{\bar{D}^{(*)0}}^2} -\sqrt{E_{l^+}^{\ast 2}-m_{l^+}^2})^2, \nonumber\\
u_{min}&=&(E^{\ast}_{\bar{D}^{(*)0}}+E^{\ast}_{l^+})^2
-(\sqrt{E_{\bar{D}^{(*)0}}^{\ast 2}-m_{\bar{D}^{(*)0}}^2}
+\sqrt{E_{l^+}^{\ast 2}-m_{l^+}^2})^2;
\end{eqnarray}
where $E^{\ast}_{\bar{D}^{(*)0}}$ and $E^{\ast}_{l^+}$ are the
energies of the charmonium state and the lepton in the rest frame of
lepton pair respectively
\begin{equation}
E^{\ast}_{\bar{D}^{(*)0}}=  {m_{\psi}^2 -m_{\bar{D}^{(*)0}}^2 -q^2
\over 2 \sqrt{q^2}}, \qquad E^{\ast}_{l^+}={\sqrt{q^2} \over 2}.
\end{equation}

Besides, the explicit form of the decay amplitude $M$ of $J/\psi \to
\bar{D}^{(*)0} l^{+} l^{-}$ is written as
\begin{eqnarray}
M_{\psi \to \bar{D}^{(*)0}l^{+} l^{-}}&=& -{G_{F} \over 4
\sqrt{2}}{\alpha_{em} \over \pi} \bigg \{\langle
\bar{D}^{(*)0}|(C_9^{eff}(\mu) \bar{u}\gamma_{\mu}(1-\gamma_5)c
-2m_c C_7^{eff}(\mu)\bar{u}i\sigma_{\mu \nu}{q^{\nu} \over
q^2}(1+\gamma_5)c )|J/\psi\rangle  \bar{l}\gamma^{\mu}l
\nonumber \\
&&+\langle \bar{D}^{(*)0}|C_{10}(\mu)
\bar{u}\gamma_{\mu}(1-\gamma_5)c |J/\psi \rangle
\bar{l}\gamma^{\mu}\gamma_5 l \bigg \}.
\end{eqnarray}
Then we can obtain the branching ratios for semi-leptonic decay of
$J/\psi \to D^{(*)-}_{d,s}$ and $\bar{D}^{(*)0}$  as
\begin{equation}
\begin{array}{ll}
{\rm{BR}}(J/\psi \to \bar{D}^{0} e^{+} e^{-})=1.14^{+0.71}_{-0.35}
\times 10^{-13}, & {\rm{BR}}(J/\psi \to \bar{D}^{0} {\mu}^{+}
{\mu}^{-})=1.08^{+0.67}_{-0.33}\times 10^{-13},
\nonumber \\
{\rm{BR}}(J/\psi \to \bar{D}^{*0} e^{+} e^{-})=6.30^{+3.61}_{-2.30}
\times 10^{-13}, & {\rm{BR}}(J/\psi \to \bar{D}^{*0} {\mu}^{+}
{\mu}^{-})=5.94^{+3.36}_{-2.15} \times 10^{-13},
\end{array}
\end{equation}
where we have combined the various uncertainties for form factors
presented in last section into the results. As can be observed, the
decay rates for the FCNC processes of $J/\psi \to \bar{D}^{(*)0}
l^{+} l^{-}$ should be very small, even including the effect from
resonances which may enhance the branching ratios considerably, if
only the SM applies.

\section{Discussions and conclusions}
\label{Discussions and conclusions}

The weak decays of $J/\psi$ meson may serve as a complementary test
of the underlying dynamics, especially the role of the FCNC in weak
decays compared with the strong and electromagnetic processes which
absolutely dominate the $J/\psi$ lifetime, even though it is very
difficult to be experimentally observed. Due to the progress of
detection facilities and techniques, it might be feasible to measure
so small branching ratios with relatively clean background and huge
database in the future. Of course, if the measurement is feasible,
it would be a better platform for determining the CKM entries
because of absence of contamination from the spectator.

Moreover, the rare weak decays of $J/\psi$ may be sensitive to the
new physics beyond the standard model. Once such weak decays were
observed with a sizable branching rate in the future colliders, it
would be a clear signal of new physics effects.

In this work, we calculate the weak decay rate of $J/\psi \to
\bar{D}^{(*)0} l^{+} l^{-}$ which is realized via FCNC-induced
processes in the framework of SM, we find that the rate is too small
to be observed in the facilities available at present. Namely, our
numerical results show that the branching ratios of such decays are
at the order of $10^{-13}$. In the calculations, we have used the
QCD sum rules and taken into account possible uncertainties coming
from both theoretical and experimental sides. Even though the
predicted branching ratios are beyond the reach of present
facilities which can be seen from a rough order estimate, a more
accurate formulation of the three point correlation function derived
in this work has theoretical significance and the technique can also
be applied to other places. In analog to some complicated
theoretical derivations which do not have immediate phenomenological
application yet, if the future experiments can provide sufficient
luminosity and accuracy, the results would be helpful.

\section*{Acknowledgements}

This work is partly supported by National Science Foundation of
China under Grant No.10745002, 10735080 and 10625525 and the special
Grant of the National Education Ministry of China.

\appendix

\section{The detailed expressions of basic functions related to the
flavor-changing neutral current processes}
\label{FCNC function}

This appendix is devoted to the collection of the basic functions
associating with flavor-changing neutral current processes, which
are taken from \cite{aliev} as
\begin{eqnarray}
C^{box}(x_q)&=&{3 \over 8}\bigg[-{1 \over x_q-1}+{x_q {\rm{ln}} x_q
\over (x_q-1)^2}\bigg], \nonumber \\
C^Z(x_q)&=&{x_q \over 4}-{3 \over 8}{1 \over x_q-1}+ {3 \over 64}{2
x_q^2 -x_q \over (x_q-1)^2}{\rm{ln}} x_q, \nonumber \\
F_1(x_q)&=&Q_q \bigg \{ \bigg[{1 \over 12} {1 \over x_q-1} + {13
\over 12}{1 \over (x_q-1)^2}-{1 \over 2 (x_q-1)^3} \bigg] x_q
\nonumber \\
&&+ \bigg[ {2 \over 3} {1 \over x_q-1} +\bigg({2 \over 3} {1 \over
(x_q-1)^2}-{5 \over 6} {1 \over (x_q-1)^3}+{1 \over 2} {1 \over
(x_q-1)^4}\bigg)x_q \bigg] {\rm{ln}} x_q\bigg \}\nonumber \\
&& -\bigg[ {7 \over 3} {1 \over x_q-1}+ {13 \over 12} {1 \over
(x_q-1)^2}-{1 \over 2} {1 \over (x_q-1)^3} \bigg] x_q \nonumber \\
&&-\bigg[ {1 \over 6} {1 \over x_q-1}- {35 \over 12} {1 \over
(x_q-1)^2}-{5 \over 6} {1 \over (x_q-1)^3} + {1 \over 2} {1 \over
(x_q-1)^4}\bigg] x_q {\rm{ln}} x_q, \nonumber \\
F_2(x_q)&=&-Q_q \bigg \{ \bigg[ -{1 \over 4} {1 \over x_q-1}+
{3\over 4} {1 \over (x_q-1)^2} +{3 \over 2} {1 \over
(x_q-1)^3}\bigg] -{3 \over 2} {x_q^2 {\rm{ln}} x_q \over (x_q-1)^4
}\bigg \} \nonumber \\
&& +\bigg[ {1 \over 2} {1 \over x_q-1}+ {9\over 4} {1 \over
(x_q-1)^2} +{3\over 2} {1 \over (x_q-1)^3}\bigg] x_q - {3 \over 4 }
{x_q^3 {\rm{ln}} x_q \over (x_q-1)^4 },
\end{eqnarray}
where $Q_q$ is the charge of down quarks $q=d,s,b$.

\section{The Wilson coefficients for $\tilde{\Pi}_{\mu \nu }$}
\label{Wilson coefficients for D0}

This appendix is devoted to the collection of visible Borel
transformed forms of  Wilson coefficients responsible for the tensor
current transition corresponding to the decay of $J/\psi$ to
$\bar{D}^{0}$ as been presented in Eq. (\ref{T1 pseudoscalar}
-\ref{T3 pseudoscalar}). It can be observed from the text that both
perturbative diagram and gluon condensate diagrams contribute to the
correlation functions non-trivially, which is distinct from the
chiral current transition remarkably. In the mathematical language,
it can be written as
\begin{eqnarray}
\tilde{f}_i=\tilde{f}_i^{pert} {\mathbf{I}}+ \tilde{f}_i^{GG}
\langle GG \rangle+O(\alpha_s)+ O({1 / m_{h}}),
\end{eqnarray}
where $\tilde{f}_i^{pert}$, $\tilde{f}_i^{GG}$ can connect with
$\tilde{\rho}^{pert}_i$ and $\tilde{\rho}^{GG}_i$ in light of the
following formulae
\begin{eqnarray}
\tilde{f}_i^{pert}&=&\int^{s_2^0}_{(m_c+m_u)^2} ds_2
\int^{s_1^0}_{s_1^L} ds_1 {\tilde{\rho}_{i}^{pert}(s_1,s_2,q^2)
\over (s_1-p_1^2) (s_2-p_2^2)} \\
\tilde{f}_i^{GG}&=&\int^{s_2^0}_{(m_c+m_u)^2} ds_2
\int^{s_1^0}_{s_1^L} ds_1 {\tilde{\rho}_{i}^{GG}(s_1,s_2,q^2) \over
(s_1-p_1^2) (s_2-p_2^2)} ,
\end{eqnarray}
or
\begin{eqnarray}
\hat{\mathcal{B}}\tilde{f}_i^{pert}= \int^{s_2^0}_{(m_c+m_u)^2} ds_2
\int^{s_1^0}_{s_1^L} ds_1 {1 \over M_1^2}   {\mathrm{e}}^{-s1/M_1^2}
{1 \over M_2^2} {\mathrm{e}}^{-s_2/M_2^2}
{\tilde{\rho}_{i}}^{pert}(s_1,s_2,q^2) \\
\hat{\mathcal{B}}\tilde{f}_i^{GG}= \int^{s_2^0}_{(m_c+m_u)^2} ds_2
\int^{s_1^0}_{s_1^L} ds_1 {1 \over M_1^2} G{\mathrm{e}}^{-s1/M_1^2}
{1 \over M_2^2} {\mathrm{e}}^{-s_2/M_2^2}
{\tilde{\rho}_{i}}^{GG}(s_1,s_2,q^2).
\end{eqnarray}
The lowest bound of $s_1$, i.e., $s_1^L$ can be determined by the
Eq. (\ref{integral region}) as
\begin{eqnarray}
s_1^L&=&- {1 \over 2 m_q^2}\bigg[m_c^4 -(2 m_q^2 +s_2
+q^2)m_c^2+m_q^2+s_2
q^2-m_q^2(s_2+q^2) \nonumber \\
&&+\sqrt{m_c^4-2(m_q^2+s_2)m_c^2+(m_q^2-s_2)^2}
\sqrt{m_c^4-2(m_q^2+q^2)m_c^2+(m_q^2-q^2)^2} \bigg], \label{s1L}
\end{eqnarray}
according to the Landau equation \cite{landau equation, S matrix}.
The detailed expressions of $\tilde{\rho}^{pert}_i$ and
$\tilde{\rho}^{GG}_i$ are given as
\begin{eqnarray}
{\tilde{\rho}}_{0}^{pert}(s_1,s_2,q^2)&=&\frac{3}{8\pi^2\lambda^{3/2}}
\{2s_1m_c^4-(4s_1m_u^2+(s_1-s_2)^2-
\lambda)m_c^2+2m_um_c\lambda+2m_u^4s_1+m_u^2((s_1-s_2)^2-\lambda)\nonumber \\
&&+(s_1+s_2)(\lambda-(s_1-s_2)^2)+q^2(2((s_1+s_2)m_c^2+s_1^2+s_2^2+s_1s_2-m_u^2(s_1+s_2))\nonumber \\
&&-(m_c^2-m_u^2+s_1+s_2)q^2)\},\nonumber
\\
{\tilde{\rho}}_{2}^{pert}(s_1,s_2,q^2)&=&\frac{1}{8\pi^2\lambda^{5/2}}
\{3s_1(\lambda^2-(2m_c^2-2m_u^2-s_1+s_2+q^2)^2\lambda -2s_1(-2m_c^2+2m_u^2+s_1-s_2-q^2)\lambda \nonumber \\
&&+2(s_1-s_2-q^2)(6s_1m_c^4+2(\lambda-3s_1(2m_u^2+s_1-s_2))m_c^2\nonumber\\
&&+s_1(6m_u^2+6(s_1-s_2)m_u^2+(s_1-s_2)^2)+s_1q^2(6m_c^2-6m_u^2-2s_1+4s_2+q^2)))\},\nonumber
\\
{\tilde{\rho}}_{4}^{pert}(s_1,s_2,q^2)&=&\frac{1}{8\pi^2\lambda^{5/2}}
\{3((2m_c(m_u-m_c)+s_1)\lambda^2-2s_1(s_2(-s_1+s_2-q^2)+(m_c^2-m_u^2)(s_1+s_2-q^2))\lambda \nonumber \\
&&-s_1(s_1-s_2-q^2)(-2m_c^2+2m_u^2+s_1-s_2-q^2)\lambda -2(s_1-s_2-q^2)((s_1+s_2)\lambda m_c^2 \nonumber \\
&&+s_1(3(s_1+s_2)m_c^4
-2(3(s_1+s_2)m_u^2+(s_1-s_2)(s_1+2s_2))m_c^2 \nonumber \\
&&+(s_1-s_2)^2s_2+3m_u^4(s_1+s_2)+2m_u^2(s_1-s_2)(s_1+2s_2))\nonumber\\
&&+q^2(-\lambda m_c^2+s_1(s_2^2+(-2m_c^2+2m_u^2+s_1)s_2+(m_c^2-m_u^2)(-3m_c^2+3m_u^2+4s_1))\nonumber\\
&&-2s_1(m_c^2-m_u^2+s_2)q^2)))\},\nonumber
\\
{\tilde{\rho}}_{5}^{pert}(s_1,s_2,q^2)&=&\frac{1}{8\pi^2\lambda^{5/2}}
\{3((m_c-m_u)\lambda(m_u s_1+m_c s_2-m_c q^2)\nonumber
\\&&+(s_1-s_2-q^2)(\lambda
m_c^2+(m_c^2-m_u^2)s_1(m_c^2-m_u^2-s_1+s_2)+s_1(m_c^2-m_u^2+s_2)q^2))\}\nonumber,
\\
{\tilde{\rho}}_{0}^{GG}(s_1,s_2,q^2)&=&\frac{64\pi^2(s_1-q^2)}{\lambda^{3/2}},\nonumber
\\
{\tilde{\rho}}_{2}^{GG}(s_1,s_2,q^2)&=&\frac{128\pi^2s_1((s_1-s_2)(s_1+2s_2)
+q^2(-2s_1+s_2+q^2))}{\lambda^{5/2}},\nonumber
\\
{\tilde{\rho}}_{4}^{GG}(s_1,s_2,q^2)&=&-\frac{64\pi^2(s_1(s_1-s_2)(s_1+5s_2)
-q^2(3s_1^2+6s_2s_1+s_2^2+q^2(-3s_1-2s_2+q^2)))}{\lambda^{5/2}},\nonumber
\\
{\tilde{\rho}}_{5}^{GG}(s_1,s_2,q^2)&=&-\frac{16\pi^2(s_1+s_2-q^2)(-s_1+s_2+q^2)}{\lambda^{3/2}}.
\end{eqnarray}

\section{The Wilson coefficients for $\tilde{\Pi}_{\mu \nu \rho}$}
\label{Wilson coefficients for D0 star}

This appendix is concentrated on Wilson coefficients relevant for
the tensor density transition of $J/\psi$ to $\bar{D}^{0}$ decay
appeared in Eq.~(\ref{tilde T1 vector}-\ref{tilde T7 vector}) after
doing the double Borel transformation. As mentioned before, both the
perturbative and gluon condensate parts are nonzero in the operator
product expansion of three-point function accounting for the tensor
operator's matrix element, which can be written as
\begin{eqnarray}
\tilde{F}_i=\tilde{F}_i^{pert} {\mathbf{I}}+ \tilde{F}_i^{GG}
\langle GG \rangle+O(\alpha_s)+ O({1 / m_{h}}).
\end{eqnarray}
The connections of $\tilde{F}_i^{pert}$ and $\tilde{F}_i^{GG}$ with
$\tilde{\rho'}^{pert}_i$, $\tilde{\rho'}^{GG}_i$ can be expressed as
\begin{eqnarray}
\tilde{F}_i^{pert}&=&\int^{s_2^0}_{(m_c+m_u)^2} ds_2
\int^{s_1^0}_{s_1^L} ds_1 {\tilde{\rho'}_{i}^{pert}(s_1,s_2,q^2)
\over (s_1-p_1^2) (s_2-p_2^2)} \\
\tilde{F}_i^{GG}&=&\int^{s_2^0}_{(m_c+m_u)^2} ds_2
\int^{s_1^0}_{s_1^L} ds_1 {\tilde{\rho'}_{i}^{GG}(s_1,s_2,q^2) \over
(s_1-p_1^2) (s_2-p_2^2)} ,
\end{eqnarray}
or
\begin{eqnarray}
\hat{\mathcal{B}}\tilde{F}_i^{pert}= \int^{s_2^0}_{(m_c+m_u)^2} ds_2
\int^{s_1^0}_{s_1^L} ds_1 {1 \over M_1^2}   {\mathrm{e}}^{-s1/M_1^2}
{1 \over M_2^2} {\mathrm{e}}^{-s_2/M_2^2}
{\tilde{\rho'}_{i}}^{pert}(s_1,s_2,q^2) \\
\hat{\mathcal{B}}\tilde{F}_i^{GG}= \int^{s_2^0}_{(m_c+m_u)^2} ds_2
\int^{s_1^0}_{s_1^L} ds_1 {1 \over M_1^2} G{\mathrm{e}}^{-s1/M_1^2}
{1 \over M_2^2} {\mathrm{e}}^{-s_2/M_2^2}
{\tilde{\rho'}_{i}}^{GG}(s_1,s_2,q^2),
\end{eqnarray}
with the lower limit of integrals $s_1^L$ defined as before.
Besides, the obvious forms of $\tilde{\rho'}^{pert}_i$,
$\tilde{\rho'}^{GG}_i$ can be displayed as
\begin{eqnarray}
{\tilde{\rho'}}_{1}^{pert}(s_1,s_2,q^2)&=&-\frac{3}{4\pi^2\lambda^{5/2}}
\{(-2(m_c+m_u)s_1(s_1-s_2)(m_c^2-m_u^2-s_1+s_2)(m_c-m_u)^2-m_c\lambda^2 \nonumber\\
&&-(2(s_1-2s_2)m_c^3+2m_u(s_2-2s_1)m_c^2-(s_1^2+s_2^2-2(m_u^2+s_1)s_2)m_c+2m_u^3s_1)\lambda \nonumber\\
&&+q^2(2(m_c-m_u)s_1(m_c^4-2(m_u^2+s_1-s_2)m_c^2+m_u^4+2m_u^2(s_1-s_2)+s_2(s_2-s_1)) \nonumber\\
&&-2m_c((m_c-m_u)m_u+s_1+s_2)\lambda+(2(m_c-m_u)s_1(m_c^2-m_u^2+s_2)+m_c\lambda)q^2))\}\nonumber
\\
{\tilde{\rho'}}_{4}^{pert}(s_1,s_2,q^2)&=&-\frac{3}{4\pi^2\lambda^{5/2}}
\{(4s_1(s_1-s_2)m_c^5+4m_us_1(s_2-s_1)m_c^4 \nonumber\\
&&+2s_1(3\lambda-(4m_u^2+3s_1-3s_2)(s_1-s_2))m_c^3+2m_us_1((4m_u^2+3s_1-3s_2)(s_1-s_2)-\lambda)m_c^2 \nonumber\\
&&+(\lambda^2-(s_2^2-4s_1s_2+3s_1(2m_u^2+s_1))\lambda+2s_1(s_1-s_2)(m_u^2+s_1-s_2)(2m_u^2+s_1-s_2))m_c \nonumber\\
&&-2m_us_1(m_u^2+s_1-s_2)((s_1-s_2)(2m_u^2+s_1-s_2)-\lambda)\nonumber\\
&&+q^2(4(m_c-m_u)s_1(m_c^4+(-2m_u^2+s_1+s_2)m_c^2+m_u^4+s_1(s_2-s_1)-m_u^2(s_1+s_2)) \nonumber\\
&&+2m_c(s_1+s_2)\lambda+(2(m_c-m_u)s_1(m_c^2-m_u^2+s_1+s_2)-m_c\lambda)q^2))\}\nonumber
\\
{\tilde{\rho'}}_{5}^{pert}(s_1,s_2,q^2)&=&-\frac{3}{4\pi^2\lambda^{5/2}}
\{((s_1-s_2)((-2m_c^3+2m_um_c^2+s_2m_c+m_us_1)\lambda \nonumber\\
&&-2(m_c-m_u)^2(m_c+m_u)s_1(m_c^2-m_u^2-s_1+s_2)) \nonumber\\
&&+q^2(2(m_c-m_u)s_1(m_c^4-2(m_u^2+s_1-s_2)m_c^2+m_s^4+2m_u^2(s_1-s_2)+s_2(s_2-s_1)) \nonumber\\
&&+(2m_c^2(m_c-m_u)-(m_c+m_u)s_1)\lambda+(2(m_c-m_u)s_1(m_c^2-m_u^2+s_2)+m_c\lambda)q^2))\}\nonumber
\\
{\tilde{\rho'}}_{7}^{pert}(s_1,s_2,q^2)&=&-\frac{3}{8\pi^2\lambda^{5/2}}
\{((m_c+m_u)((m_c-m_u)^2-s_2)\lambda-(m_c+m_u)(s_2(-s_1+s_2-q^2) \nonumber\\
&&+(m_c^2-m_u^2)(s_1+s_2-q^2))(s_1-s_2+q^2) \nonumber\\
&&+2(m_c-m_u)(\lambda m_c^2+(m_c^2-m_u^2)
s_1(m_c^2-m_u^2-s_1+s_2)+s_1(m_c^2-m_u^2+s_2)q^2))\}\nonumber
\\
{\tilde{\rho'}}_{9}^{pert}(s_1,s_2,q^2)&=&\frac{3}{8\pi^2\lambda^{3/2}}
\{((m_c-m_u)(s_2(-s_1+s_2-q^2)+(m_c^2-m_u^2)(s_1+s_2-q^2))(s_1-s_2+q^2) \nonumber \\
&&+\lambda(m_c^3-m_um_c^2-m_u^2m_c-q^2m_c+m_u(m_u^2+s_1-s_2)))\},\nonumber
\\
{\tilde{\rho'}}_{11}^{pert}(s_1,s_2,q^2)&=&\frac{3}{8\pi^2\lambda^{3/2}}
\{(m_c+m_u)s_1(\lambda+(2(m_c-m_u)^2+s_1-s_2-q^2)(2m_c^2-2m_u^2-s_1+s_2+q^2))\},\nonumber
\\
{\tilde{\rho'}}_{15}^{pert}(s_1,s_2,q^2)&=&-\frac{3}{4\pi^2\lambda^{5/2}}
\{(m_c\lambda^2-(m_c-m_u)(s_2(-s_1+s_2-q^2)+(m_c^2-m_u^2)(s_1+s_2-q^2))\lambda \nonumber\\
&&-(m_c+m_s)(s_2(-s_1+s_2-q^2)+(m_c^2-m_u^2)(s_1+s_2-q^2))\lambda \nonumber\\
&&+(m_c-m_u)s_1(2m_c^2-2m_u^2-s_1+s_2+q^2)\lambda \nonumber\\
&&-2(m_c-m_u)((s_1+s_2)\lambda
m_c^2+s_1(3(s_1+s_2)m_c^4-2(3(s_1+s_2)m_u^2+(s_1-s_2)(s_1+2s_2))m_c^2 \nonumber\\
&&+(s_1-s_2)^2s_2+3m_u^4(s_1+s_2)+2m_u^2(s_1-s_2)(s_1+2s_2)) \nonumber\\
&&+q^2(-\lambda
m_c^2+s_1(s_2^2+(-2m_c^2+2m_u^2+s_1)s_2+(m_c^2-m_u^2)(-3m_c^2+3m_u^2+4s_1))\nonumber\\
&&-2s_1(m_c^2-m_u^2+s_2)q^2)))\} \nonumber
\\
{\tilde{\rho'}}_{1}^{GG}(s_1,s_2,q^2)&=&\frac{1}{\lambda^{5/2}}
\{128(m_c-m_u)\pi^2(q^4-(2s_1+s_2)q^2+s_1(s_1-s_2))\}\nonumber
\\
{\tilde{\rho'}}_{4}^{GG}(s_1,s_2,q^2)&=&-\frac{1}{\lambda^{5/2}}
\{256(m_c-m_u)\pi^2s_1(s_1-s_2+2q^2)\}\nonumber
\\
{\tilde{\rho'}}_{5}^{GG}(s_1,s_2,q^2)&=&\frac{1}{\lambda^{5/2}}
\{128(m_c-m_u)\pi^2(s_1-s_2-q^2)(2s_1-s_2+q^2)\}\nonumber
\\
{\tilde{\rho'}}_{7}^{GG}(s_1,s_2,q^2)&=&-\frac{1}{\lambda^{5/2}}
\{64(m_c-m_u)\pi^2(s_1(s_1-s_2^2)^2-q^2(3s_1^2+12s_1s_2+s_2^2+q^2(-3s_1-2s_2+q^2)))\}\nonumber
\\
{\tilde{\rho'}}_{9}^{GG}(s_1,s_2,q^2)&=&\frac{1}{\lambda^{5/2}}
\{64(m_c-m_u)\pi^2(s_1-s_2-q^2)((s_1-s_2)(2s_1+s_2)-q^2(s_1-2s_2+q^2))\}\nonumber
\\
{\tilde{\rho'}}_{11}^{GG}(s_1,s_2,q^2)&=&\frac{1}{\lambda^{5/2}}
\{128(m_c-m_u)\pi^2s_1((s_1-s_2)^2+4q^4-5(s_1+s_2)q^2)\}\nonumber
\\
{\tilde{\rho'}}_{15}^{GG}(s_1,s_2,q^2)&=&-\frac{1}{\lambda^{7/2}}
\{128(m_c-m_u)\pi^2((s_1^2+10s_1s_2+s_2^2)(s_1-s_2)^2\nonumber\\
&&+q^2(q^2(-3s_1^2-50s_1s_2-3s_2^2-2q^4+5(s_1+s_2)q^2)-(s_1+s_2)(s_1^2-38s_1s_2+s_2^2)))\}.
\end{eqnarray}

\end{document}